\documentclass[a4paper,UKenglish,cleveref, autoref, thm-restate]{lipics-v2021}

\hideLIPIcs

\nolinenumbers

\usepackage{proof}
\usepackage{mathtools}
\usepackage{stmaryrd}
\usepackage[capitalise]{cleveref}
\usepackage[all]{xy}
\usepackage{bbm}
\usepackage{nicefrac}
\usepackage{enumitem}
\usepackage{multicol}
\usepackage{xspace}
\crefname{theorem}{Theorem}{Theorems}
\crefname{lemma}{Lemma}{Lemmata}
\crefname{remark}{Remark}{Remarks}
\crefname{example}{Example}{Examples}
\crefname{proposition}{Proposition}{Propositions}
\crefname{definition}{Definition}{Definitions}
\crefname{section}{\S}{\S\S}
\Crefname{section}{Section}{Sections}

\newcommand{\name}{conditionally branching\xspace}
\newcommand{\Name}{Conditionally branching\xspace}	
\newcommand{\ifte}{\texttt{if-then-else}\xspace}
\newcommand{\resp}{resp.\@\xspace}
\newcommand{\cf}{\textit{cf.}\@\xspace}
\newcommand{\ie}{\textit{i.e.}\@\xspace}
\newcommand{\eg}{\textit{e.g.}\@\xspace}

\newcommand{\wrt}{w.r.t.\@\xspace}

\newcommand{\defeq}{\triangleq}
\newcommand{\inv}{^{-1}}
\newcommand{\inj}{\mathrm{in}}
\newcommand{\id}{\mathrm{id}}
\newcommand{\sem}[1]{\llbracket #1 \rrbracket}
\newcommand{\R}{\mathbb{R}}
\newcommand{\C}{\mathbb{C}}

\newcommand{\one}{\mathbbm{1}}
\newcommand{\bra}[1]{\left\langle #1\right\rvert}
\newcommand{\ket}[1]{\left\lvert #1\right\rangle}
\DeclarePairedDelimiter\abs{\lvert}{\rvert}%
\DeclarePairedDelimiter\norm{\lVert}{\rVert}%
\newcommand{\cat}[1][C]{\mathscr{#1}}
\newcommand{\catFont}[1]{\mathsf{#1}}
\newcommand{\Set}{\catFont{Set}}
\newcommand{\oCPO}{\omega\catFont{CPO}}
\newcommand{\oCPOs}{\omega\catFont{CPO}_\bot}
\newcommand{\Vect}{\catFont{Vec}}

\newcommand{\Rel}{\catFont{Rel}}
\newcommand{\Ban}{\catFont{Ban}}
\newcommand{\MS}{\catFont{Meas}}
\newcommand{\Hilb}{\catFont{Hilb}}
\newcommand{\Stoch}{\catFont{Stoch}}
\newcommand{\Kl}[1]{\cat_{#1}}
\newcommand{\CPM}{\catFont{CPM}}
\newcommand{\CPMp}{\CPM_\oplus}
\newcommand{\Meas}{\mathcal{M}}

\newcommand{\Id}{\mathsf{Id}}
\newcommand{\ldistr}{\mathrm{ldistr}}
\newcommand{\rdistr}{\mathrm{rdistr}}
\DeclareMathOperator{\bang}{!}

\newcommand{\dg}{^\dagger}
\newcommand{\progfont}[1]{\mathtt{#1}}
\newcommand{\typefont}[1]{\progfont{#1}}
\newcommand{\type}[1][T]{\typefont{#1}}
\newcommand{\ite}[3]{\progfont{if}~#1~\progfont{then}~#2~\progfont{else}~#3}

\newcommand{\ptrue}{\mathtt{tt}}

\newcommand{\pin}{\progfont{in}}
\newcommand{\caseof}[1]{\progfont{case~}#1\progfont{~of~}}
\newcommand{\pnot}{\mathop{\progfont{not}}}

\newcommand{\qubit}{\progfont{qubit}}
\newcommand{\bit}{\progfont{bit}}
\newcommand{\meas}{\progfont{meas}}
\newcommand{\cp}{\mathrm{copy}}
\newcommand{\del}{\mathrm{del}}
\DeclareMathOperator{\strue}{\mathrm{true}}
\DeclareMathOperator{\sfalse}{\mathrm{false}}
\DeclareMathOperator{\snot}{\mathrm{not}}
\DeclareMathOperator{\cho}{\mathtt{C}}

\title{On the semantics of conditional branching.} 

\author{Fredrik Dahlqvist}{Queen Mary University London}{f.dahlqvist@qmul.ac.uk}{https://orcid.org/ 0000-0003-2555-0490}{}

\authorrunning{F. Dahlqvist} 

\Copyright{Fredrik Dahlqvist} 



\category{} 

\relatedversion{} 

\EventEditors{John Q. Open and Joan R. Access}
\EventNoEds{2}
\EventLongTitle{42nd Conference on Very Important Topics (CVIT 2016)}
\EventShortTitle{CVIT 2016}
\EventAcronym{CVIT}
\EventYear{2016}
\EventDate{December 24--27, 2016}
\EventLocation{Little Whinging, United Kingdom}
\EventLogo{}
\SeriesVolume{42}
\ArticleNo{23}

\begin{document}

\maketitle

\begin{abstract}
Motivated by the non-standard interpretations of \ifte in probabilistic and quantum programs we review the different semantics of conditional branching and propose a unifying categorical structure for the interpretation of \ifte. We show that many semantically important categories possess this structure, covering \textit{inter alia} classical, probabilistic, quantum or non-deterministic computations. We examine algebraic structures related to and generated by this categorical structure, in particular on collections of predicates. In this setting we show that the skew-associativity rules of GKATs and barycentric algebras are two instances of a common skew-associativity rule.  
\end{abstract}

\section{Introduction}\label{sec:intro}

\textbf{Motivation.}
The ideas and results presented in this work originate from a simple observation about the semantics of probabilistic programs. If one follows the classic semantics proposed by Kozen either in terms of forward state (\ie measure) transformer~\cite{K79c}, or in terms of backwards predicate (\ie function) transformer~\cite{K85a}, one inevitably lands in a semantic world of vector spaces and operators. This naturally leads to a \emph{linear} perspective on probabilistic programs where states cannot be copied, or at least not in a way that is compatible with higher order computations~\cite{dahlqvist2019semantics}. In other words, randomness is seen as a resource.

However, there is one fundamental construct with respect to which randomness does \emph{not} behave as a resource: conditional branching, that is to say \ifte statements. Consider for example the following simple imperative program
\begin{lstlisting}
if x>0 then x = x+1 else x = x-1
\end{lstlisting}
Following~\cite{K79c}, this program is interpreted as a linear operator on the Banach space of input distributions on $\mathtt{x}$. Given such an input state $\mu$, the semantics of the program will:
\begin{itemize}
	\item[(i)]  \emph{split} $\mu$ into two subdistributions, one satisfying the guard and the other its complement,
	\item[(ii)] apply the \texttt{if} branch ($\mathtt{+1}$) to the first subdistribution,  the \texttt{else} branch ($\mathtt{-1}$) to the second,
	\item[(iii)]finally, \emph{recombine} the two branches (via pointwise addition) to produce the output.
\end{itemize} 
The remarkable fact is that, despite being a fundamentally linear semantics, \emph{the initial state $\mu$ is not consumed by the evaluation of the guard} $\mathtt{x>0}$. In fact, the guard is not evaluated at all in the sense that no boolean value or even convex combination thereof is produced. The splitting-recombining mechanism provides a linear alternative to guard evaluation that allows each branch to access the (relevant part of the) input state.

The situation is very similar in the quantum case, although it is complicated by the existence of two fundamentally different notions of conditional branching. The first is \emph{coherent} or \emph{quantum} branching, where the guard is itself a quantum system and no measurement is performed. A canonical example is the CNOT gate, where the control qubit determines which branch is taken while remaining available and un-measured. Semantically, this type of branching can be understood as an instance of the mechanism we have just described: the quantum state is split and then recombined as a superposition of two control paths (the \texttt{if} and \texttt{else} branches), each transformed independently. The other notion is \emph{measurement-based} branching, where the guard is evaluated by measuring a quantum system and the branch is selected by the resulting classical outcome. Here the measurement irreversibly changes the observed qubits and may alter any entanglement between them and the rest of the system. 

The state-splitting mechanism of probabilistic and coherent control is different both from the classical interpretation of conditional branching~\cite{gibbons1997conditionals} where the state/context is copied, with one copy used to evaluate the guard and the other to continue the computation, and from the linear interpretation of conditional branching where evaluating the guard consumes the state/context making it unavailable for the rest of the computation (as is the case in measurement-based quantum control~\cite{pagani2014applying}). There are thus at least three incomparable ways of interpreting \ifte statements, which raises the following questions:
\begin{enumerate}
	\item \textit{Can we reconcile these different interpretations of conditional branching?} 
	\item \textit{More generally, what is the `minimal' structure needed to interpret \emph{\ifte}?}
	\item \textit{What program equivalences can we expect of a general interpretation of conditionals?}
\end{enumerate}
Answering these questions is about a lot more than clarifying an interesting quirk of probabilistic program semantics, \ifte statements are fundamental to the practice of programming	and we should endeavour to interpret and understand them in the widest possible context. Concretely, this means that given any programming paradigm, say probabilistic hybrid computations with exceptions, we should be able to work out (a) what (if any) interpretations of conditional branching it can support, and (b) what kind of program equivalences can be shown for these interpretations. For instance does $\progfont{if~b~then~p~else~q}\equiv \progfont{if~not~b~then~q~else~p}$ hold? What does $\progfont{not}$ even mean? 

The objective of this work is to provide tools for answering these questions. As a short slogan: we aim to provide a semantic structure that does for conditional branching what traced monoidal categories did for recursion and iteration. 

\smallskip
\textbf{Contributions and structure of the paper.}
In \cref{sec:wild} we review the various semantics of conditional branching that can be found `in the wild': the classical interpretation via copying and distributing (\cref{subsec:classical}), the linear and mixed interpretations (\cref{subsec:linear}), the probabilistic `state-splitting' approach (\cref{subsec:splitting}), and coherent control (\cref{subsec:coherent}).
In \cref{sec:def} we introduce and justify the notion of \emph{\name category} as a semantic structure in which \ifte statements can be interpreted in a way that follows (some of) the standard axioms governing the behaviour of \ifte statements. 
In \cref{sec:examples} we provide examples of \name categories including domain-theoretic models of call-by-value computations, Markov categories~\cite{fritz2020synthetic}, models of effectful computations~\cite{moggi1991notions}, and the motivating vector-space models for which we present a very general state-splitting result (\cref{subsec:vec}).
Finally, in \cref{sec:structure} we will see that a lot of structural mileage can be obtained from the humble coproduct. We study the structure of the set of \emph{branching morphisms}, \ie morphisms of the shape $X\to X+X$ in \cref{subsec:structure_bm} and of the set of \emph{tests},  \ie morphisms of the shape $X\to 1+1$ in \cref{subsec:structure_tests}.	
Finally, in \cref{sec:skew}	we examine the skew-associativity rule associated with
one of the more complex behaviours of conditional branching that sometimes allows us to re-write 
nested \ifte statements. 
We show that the skew-associativity axiom of Guarded Kleene Algebras with Tests~\cite{SFHKKS20a} and the seemingly unrelated skew-associativity axiom of barycentric algebras~\cite{mardare2016quantitative} are two instances of a general skew-associativity rule. 

\smallskip
\textbf{Notation and terminology.}
We will be using a generic typed functional language and consider typing judgements
$\Gamma \vdash s:\type$ interpreted as morphisms $\sem{s}:\sem{\Gamma} \to\sem{\type}$ in a category $\cat$. We will be concerned with such a small language fragment (just what is needed to talk about \ifte) that we choose not to specify a full syntax and semantics but rather introduce constructs if and when needed. 
We assume that the type grammar contains a unit type 1 and coproduct types  $\type[S]+\type$. Term formation thus includes the typing rules:
\begin{align}
	\small{
	\hspace{-14pt}
	\infer{\Gamma \vdash \pin(1,s):\type[S]+\type}{\Gamma\vdash s:\type[S]}
	\hspace{13pt}
	\infer{\Gamma \vdash \pin(2,s):\type[S]+\type}{\Gamma\vdash s:\type}
	\hspace{13pt}
	\infer{\Gamma,\Delta \vdash \caseof{s}(\pin(1,x)\Rightarrow u, \pin(2,y)\Rightarrow v):\type}{\Gamma\vdash s:\type[S]_1+\type[S]_2 & x:\type[S]_1,\Delta \vdash u:\type&  y:\type[S]_2,\Delta\vdash v:\type }}\label{rule:pattern}
\end{align}
The pattern-matching rule above is \emph{linear}: contexts are consumed by computations. We will not assume that the categories in which programs are interpreted are \emph{Cartesian}, we will only assume that they are \emph{monoidal}, as is the case for vector-space models of probabilistic and quantum computations, as well as many examples of effectful computations. However, guided by the probabilistic and quantum examples given in the introduction, this paper will revolve around understanding the semantics of the following \ifte-rule:
\begin{align}
	\small{
	\infer{\Gamma, \Delta\vdash \ite{p}{s}{t}:\type}
	{\Gamma\vdash p:(1+1) & \Gamma, \Delta \vdash s:\type& \Gamma, \Delta\vdash t:\type}}\label{rule:cond_vanilla}
\end{align}
where the context $\Gamma$ is now used in a \emph{non-linear} manner: the program $p$ does \emph{not} consume the context in \eqref{rule:cond_vanilla}, it is still available to $s$ and $t$. In other words, we will \emph{not} view $\ite{p}{s}{t}$ as syntactic sugar for instances of the pattern-matching rule \eqref{rule:pattern} of the shape $\Gamma,\Delta \vdash\caseof{p}(\pin(1,x)\Rightarrow s, \pin(2,x)\Rightarrow t)$ with a guard $\Gamma\vdash p: 1+1$. Note that this context-consuming type of branching will nevertheless always be available to us \emph{as well}, by virtue of the existence of \eqref{rule:pattern}. 
We will call programs of type $1+1$ \emph{predicates}~\cite{jacobs2015new} or \emph{tests}~\cite{kozen1997kleene}. In the context of a program $ \ite{p}{s}{t}$, the predicate $p$ will be referred to as the \emph{guard}. 

We assume basic knowledge of category theory and use the standard notations $\pi_1, \pi_2$ and $\inj_1, \inj_2$ for the first and second projection and coprojections respectively,  $\langle f,g\rangle$ and $[f,g]$ for the pairing and copairing of morphisms, $f\times g\defeq \langle f\circ \pi_1, g\circ \pi_2\rangle$ and $f+g\defeq [\inj_1\circ f, \inj_2\circ g]$, and finally $\Delta_X\defeq\langle\id_X,\id_X\rangle: X\to X\times X$ and $\nabla_X\defeq [\id_X, \id_X]: X+X\to X$ for the diagonal and codiagonal morphisms respectively.		

\section{Semantics of \texttt{if-then-else} in the wild}\label{sec:wild}

\subsection{Classical branching by copying and distributing}\label{subsec:classical}

\textbf{Vanilla flavour.}
The standard way of interpreting the \ifte rule \eqref{rule:cond_vanilla} involves copying the context, using one copy to evaluate the guard and the other to continue the computation. 
This can be formalised in any \emph{distributive category}~\cite{gibbons1997conditionals}, that is to say in any category $\cat$ with finite products and coproducts such that for any $\cat$-objects $A,B,C$ the canonical morphism
\[
[\id_A \times \inj_1, \id_A\times\inj_2]: (A\times B)+(A\times C)\to A\times(B+C)
\]
is an isomorphism. We write $\ldistr_{A,B,C}: A\times(B+C)\to(A\times B)+(A\times C)$ for its inverse. Similarly, we will write $\rdistr_{A,B,C}: (B+C)\times A \to(B\times A)+(C\times A)$ for the right-distributivity isomorphism.
In the case that interests us the most $B=C=1$ and we will therefore simply write $\ldistr_{A}: A\times(1+1)\to(A\times 1)+(A\times 1)\simeq A+A$.
In this non-linear/Cartesian context \eqref{rule:cond_vanilla} simplifies  to
\begin{align}
	\small{
	\infer{\Gamma\vdash \ite{p}{s}{t}:\type}{\Gamma\vdash p:(1+1) & \Gamma \vdash s:\type& \Gamma\vdash t:\type}\label{rule:cond_distrib}
	}
\end{align}
Given a distributive category $\cat$, the conclusion of the typing rule \eqref{rule:cond_distrib} is interpreted as
{
	\begin{align}
		\small
		\xymatrix@C=10ex{
			\sem{\Gamma}\ar[r]^-{\Delta_{\sem{\Gamma}}} & \sem{\Gamma\times\Gamma}\ar[r]^-{\id\times \sem{p}} 
			& 
			\sem{\Gamma}\times (1+1)\ar[r]^-{\ldistr_{\sem{\Gamma}}} & \sem{\Gamma+\Gamma}\ar[r]^-{\left[\sem{s},\sem{t}\right]} & \sem{\type}
		}
	\end{align}
}Whilst this is a natural and intuitive semantics, many categories that are important for the semantics of programs fail to be distributive.

\begin{example}\label[example]{ex:Rel}
	Non-deterministic computations are often interpreted in the Kleisli category of the powerset monad on $\Set$, which is equivalent to the category $\Rel$ of sets and relations. In $\Rel$, products and coproducts are both given by disjoint union~\cite[\S 5.4.7]{barr1990category} and the category is therefore not distributive: $1\times (1+1)$ has three elements whereas $(1\times 1)+(1\times 1)$ has four.
\end{example}

\begin{example}\label[example]{ex:CPO}
	Another important example is the category $\oCPO$ of pointed $\omega$-complete partial order, or $\omega$-cpos, and monotone maps. This category does not have coproducts~\cite{smyth1982category} and in particular cannot be distributive. If we restrict ourselves to the sub-category $\oCPOs$ of $\omega$-cpos and \emph{strict} morphisms,  
	then coproducts do exist and are given by the carriers
	$
	A+B = (A\setminus\{\bot_A\})+(B\setminus\{\bot_B\})+\{\bot\}
	$
	with the order inherited from $A,B$ (sometimes known as the coalesced sum).  Products are constructed by taking the Cartesian product of the carriers together with the pointwise order. However, $\oCPOs$ is not distributive as a cardinality argument shows: if we consider the chain $\mathbbm{2}$ then $\mathbbm{2}\times (\mathbbm{2}+\mathbbm{2})$ has $2\times ((2-1)+(2-1)+1)=6$ elements but $(\mathbbm{2}\times\mathbbm{2})+(\mathbbm{2}\times\mathbbm{2})$ has $((2\times 2)-1)+((2\times 2)-1)+1=7$ elements.
\end{example}

\begin{example}\label[example]{ex:Vect}
	Real-weighted (including probabilistic) computations can be interpreted in the category $\Vect$ of vector spaces and linear maps. $\Vect$ is not distributive for a reason that is similar to $\Rel$, namely because the coproduct is given by the direct sum which is isomorphic to the Cartesian product in finite dimensions.
\end{example}
We show in \cref{sec:examples} that these non-distributive examples do support a semantics for \ifte.

\textbf{Monadic flavour.}
The copy-and-distribute approach can be extended straightforwardly to effectful computations \`{a} la Moggi \cite{moggi1991notions}. This is done in~\cite{jacobs2015new,jacobs2016affine} which is probably the line of research that is closest to this work, although it focuses exclusively on Cartesian categories. 		
Given a distributive category $\cat$ and a \emph{strong} monad $T:\cat\to\cat$, we can consider predicates in the Kleisli category of $T$, \ie $\cat$-morphisms $p:X\to T(1+1)$. Given such a predicate,~\cite{jacobs2015new,jacobs2016affine} defines the associated \emph{instrument} as the morphism $\mathrm{instr}_p: X\to T(X+X)$ defined as
\[
\xymatrix@C=10ex{
	X\ar[r]^-{\langle \id,p\rangle}
	& 
	X\times T(1+1) \ar[r]^{\mathrm{str}}
	&
	T(X\times (1+1))\ar[r]^-{T\ldistr}
	& 
	T(X+X)
}
\]
With this special morphism (that we generalise in \cref{sec:def}), the semantics of \eqref{rule:cond_distrib}  is given by $\sem{\ite{p}{s}{t}}\defeq \big[\sem{s},\sem{t}\big]\circ \mathrm{instr}_{\sem{p}}$ (where composition is taken in the Kleisli category). We return to this construction in more detail in \cref{subsec:namecopy}.

\subsection{Linear and mixed branching}\label{subsec:linear}

\textbf{The linear case.} In the context of a linear language equipped with the injections and pattern-matching rules \eqref{rule:pattern}, \ifte is often understood in the context-consuming sense described in \cref{sec:intro}, that is to say as syntactic sugar for pattern-matching (\eg \cite{pagani2014applying,staton2017commutative}):
\begin{align}
	\hspace{-9pt}\ite{p}{s}{t}\hspace{-1pt}\defeq\hspace{-1pt}\caseof{p} \hspace{-1pt}(\pin(1,x)\hspace{-1pt}\Rightarrow\hspace{-1pt} s, \hspace{-1pt}\pin(2,x)\hspace{-1pt}\Rightarrow\hspace{-1pt} t)\label{eq:badifte}
\end{align}
One can interpret \eqref{eq:badifte} in any \emph{distributive monoidal} category $(\cat,\otimes,I, +)$, \ie a monoidal category that has coproducts and isomorphisms $\ldistr_{A,B,C}: A\otimes (B+C)\hspace{-1pt}\to\hspace{-1pt} (A\otimes B) + (A\otimes C)$ and $\rdistr_{A,B,C}: (B+C)\otimes A\hspace{-1pt}\to\hspace{-1pt} (B\otimes A)+(C\otimes A)$ via:
\begin{align}\label{eq:coproduct_semantics}
	\hspace{-9pt}\small{
	\xymatrix@C=10ex@R=3ex{
		\sem{\Gamma}\otimes\sem{\Delta}\ar[r]_-{\sem{p}\otimes \id}\ar@/^1.4pc/[rrr]^-{\sem{\ite{p}{s}{t}}} 
		&
		 \left(\sem{\type[S]_1}+\sem{\type[S]_2} \right)\otimes\sem{\Delta}\ar[r]_-{\rdistr}
		&
		\sem{\Delta}\otimes \sem{\type[S]_1}+\sem{\Delta}\otimes \sem{\type[S]_2}\ar[r]_-{\big[\sem{s},\sem{t}\big]}
		&
		\sem{\type}
	}}
\end{align}
Since \eqref{eq:coproduct_semantics} interprets an instance of \eqref{rule:pattern}, the variables used by the programs $s,t$ are not available to the guard $p$. This is arguably not how \ifte statements are intuitively understood or used in practice. Indeed, the point of conditional branching is typically to apply different procedures to variables depending on a predicate on these \emph{same} variables. Thus whilst we acknowledge that \eqref{eq:badifte} provides a possible interpretation for \ifte statements -- and as discussed in the introduction this interpretation will always be available to us -- we will be concerned with the interpretation of the non-linear branching rule \eqref{rule:cond_vanilla}. 

\textbf{The mixed case.} There exists a half-way house between the non-linear copy-and-distribute mechanism and the purely linear one that is used when both non-linear (duplicable) and linear (non-duplicable) types coexist. Formally, consider a distributive monoidal category $(\cat, \otimes , I, +)$ that is equipped with an exponential comonad $!$ satisfying the usual requirements and defining copiable types $!A$.
The rule \eqref{rule:cond_vanilla} can then be modified into:
\begin{align}\label{rule:cond_linear}
	\small{
		\infer{!\Gamma, \Delta, E\vdash \ite{p}{s}{t}:\type}{!\Gamma, \Delta\vdash p:(1+1) &!\Gamma, E\vdash s:\type& !\Gamma, E\vdash t:\type}}
\end{align}
and be interpreted via the exponential's comultiplication (\ie copying map) $\delta: \bang \to \bang\otimes\bang$ as:
\begin{align*}
	\small{
	\xymatrix@C=10ex@R=4ex{
		!\sem{\Gamma}\otimes\sem{\Delta}\otimes\sem{E}\ar[d]_-{\delta\otimes\id\otimes\id} \ar[rr]^-{\sem{\ite{p}{s}{t}}}
		& &	\sem{\type} 
		\\
		!\sem{\Gamma}\otimes!\sem{\Gamma}\otimes \sem{\Delta}\otimes\sem{E}\ar[r]^-{\id\otimes \sem{p}\otimes \id}  &
		!\sem{\Gamma}\otimes (1+1)\otimes \sem{E}\ar[r]^-{\ldistr\circ \rdistr} 
		&  !\sem{\Gamma}\otimes\sem{E}+!\sem{\Gamma}\otimes\sem{E}\ar[u]_-{[s,t]} 			
	}}
	\end{align*}

\begin{example}\label{ex:meas_based_ifte}
	This is the semantics of the \emph{measurement-based} branching of quantum computing, as defined for the quantum lambda calculus of~\cite{selinger2006lambda, selinger2009quantum,valiron2022semantics} and interpreted in~\cite{pagani2014applying} and~\cite{valiron2022semantics} (which specifically discusses branching). The context of the guard $p$ consists of classical data that can be copied ($!\Gamma$) and quantum data ($\Delta$) that cannot. In this model of measurement-based control, the quantum part of the input state gets collapsed by the measurement (\ie by evaluating the guard) and is lost; only the result of the evaluation of the guard (which has type 1+1) is available to continue the computation. 
	
	Concretely, let us assume ground types $\qubit$ and $\bit\defeq1+1$ and built-in functions $\meas: \qubit\to\bit$ and $\progfont{Z}: \qubit\to\qubit$ interpreted in the category $\CPMp$ whose objects are direct sums of finite-dimensional Hilbert spaces and whose morphisms are completely positive linear maps in the sense of \cite[\S 3.2]{selinger2004towards}.  We can then consider the semantics of the instance of \eqref{rule:cond_linear} (feed-forward correction) on two qubits \texttt{x,y} that measures the state of \texttt{x}, leaves \texttt{y} unchanged if  $\ket{0}$ is measured, or applies the \texttt{Z} gate to $y$ if $\ket{1}$ is measured:
		\[
		\small{
				\infer{x,y:\qubit\vdash \ite{\meas(x)}{y}{\progfont{Z}y}:\qubit}
				{x:\qubit\vdash\meas(x):\bit & y:\qubit\vdash y:\qubit & y:\qubit\vdash  \progfont{Z}y:\qubit}
			}
		\]
The semantics of this program is given by \eqref{eq:coproduct_semantics} which yields for an input density (block) matrix
\begin{align*}
	\sem{\ite{\meas(x)}{y}{\progfont{Z}y}}\begin{pmatrix}A & B \\ C& D\end{pmatrix}&\defeq \big[ \id, \sem{\progfont{Z}}\big]\circ (\sem{\meas}\otimes \id) \begin{pmatrix}A & B \\ C& D\end{pmatrix}
	\\
	&=\big[ \id, \sem{\progfont{Z}}\big] (A,D) = A + ZDZ\dg
\end{align*}
Note the disappearance of the entanglement (represented by the blocks $B,C$) due to the measurement of $x$.
We will see in \cref{subsec:vec} that the non-linear \ifte rule \eqref{rule:cond_vanilla} can alternatively be used and satisfactorily interpreted to describe measurement-based control.
\end{example}

\subsection{Branching by splitting and merging}\label{subsec:splitting}
The motivating example presented in the introduction was that of probabilistic programs, a computational paradigm which is linear (at least in its vector-space interpretation) yet comes with a meaningful mechanism for interpreting \eqref{rule:cond_vanilla}. 

Following~\cite{K79c}, we sketch parts of the semantics for \emph{first-order} probabilistic programs in the category $\Ban$ of Banach spaces (over $\R$) and linear contractions. This category is monoidal \wrt to the projective tensor product (which we will not need here) and has $I=\R$ as unit object.  The coproduct in this category is given by the direct sum equipped with the $\ell_1$-norm, thus $I+I$ is the vector space $\R^2$ equipped with the norm $\norm{(u,v)}_1=\abs{u}+\abs{v}$, and $\Ban$ is distributive monoidal for these two operations.
Each ground type $\type$ is first associated with an object $X(\type)$ in the category $\MS$ of measurable spaces and measurable functions. For instance the unit type $1$ is associated with the singleton space $X(1)=\{\ast\}$ and the numeric type $\typefont{Real}$ is associated with $X(\typefont{Real})=\R$, with the usual Borel $\sigma$-algebras. Given a measurable space $X$ we will write $\Meas X$ for the Banach space of finite measures on $X$. For a context $\Gamma=x_1:\type_1,\ldots, x_n:\type_n$, a judgement $\Gamma\vdash t: \type$ is interpreted as a linear operator
\[
\sem{t}:\sem{\Gamma}\defeq \Meas \left(X(\type_1)\times \ldots\times X(\type_n)\right)\to \sem{\type}\defeq \Meas X(\type)
\] 
The linear operator $\sem{t}$ transforms a joint probability distribution on the input variables of the context into an output distribution on the output type. A \emph{deterministic} predicate $\Gamma\vdash p: 1+1$ is a predicate whose interpretation is of the shape
\[
\sem{p}=f_\ast :   \Meas \left(X(\type_1)\times \ldots\times X(\type_n)\right) \to \Meas X(1+1)\simeq \R^2
\]
where $f_\ast$ is the pushforward  associated with a measurable map $f: X(\type_1)\times\ldots\times X(\type_n)\to X(1+1)=2$. From $p$ we can define the  \emph{linear} `decomposition' operator splitting a distribution into a pair of subdistributions:
\begin{align*}
	\widehat{\sem{p}}: \sem{\Gamma} \to \sem{\Gamma}\oplus \sem{\Gamma} \quad \mu \mapsto \left(\mu(-\cap f\inv(0)), \mu(-\cap f\inv(1))\right).
\end{align*}
This operator $\widehat{\sem{p}}$ allows us to interpret the typing judgement $ \sem{\Gamma\vdash \ite{p}{s}{t}:\type}$ as
\begin{align}\label{eq:probite}
	\xymatrix@C=8ex{
		\sem{\Gamma}\ar[r]^-{\widehat{\sem{p}}} & \sem{\Gamma}\oplus \sem{\Gamma}\ar[r]^-{\big[\sem{s},\sem{t}\big]} & \sem{\type}
	}
\end{align}
That this non-context-consuming branching mechanism can coexist with a linear system is remarkable and rests on the ability to define the splitting operator $\widehat{\sem{p}}$ as a contractive linear operator. 
We will show in \cref{subsec:vec} that this construction can be extended to arbitrary guards for a very large class of input spaces $\sem{\Gamma}$.

\subsection{Coherent branching in quantum programs}\label{subsec:coherent} 
Our last example of conditional branching found in the wild is coherent/superposed control, a type of branching found in quantum computation, for example in the CNOT gate and described as the `elusive quantum control' in~\cite[\S 4.2]{valiron2022semantics} as it is harder to account for both type-theoretically and semantically. We follow the formalism for QML~\cite{altenkirch2005functional,altenkirch2007algebra} and refer the reader to \cite{valiron2022semantics} for a detailed discussion. In QML the typing rule for coherent control is
\begin{align}\label{rule:QML_coherent}
	\small{
		\infer{\Gamma, \Delta \vdash \ite{\hspace{-4pt}\prescript{\circ}{}{\hspace{3pt}p}}{s}{t}:\type}{\Gamma\vdash p:\qubit &\Delta\vdash s:\type& \Delta\vdash t:\type}}
\end{align}
where $\mathtt{if}^\circ$ is meant to distinguish superposed control from measurement-based control and we have introduced the self-explanatory type $\qubit$. The semantics of this rule assumes the existence of products, of a natural transformation $\delta_{A,B}: A\otimes B\to A\times B$, and for any pair $f,g: A\to B$ a mechanism for combining them into a morphism $(f\mid g):(1+1)\times A\to B$ selecting $f$ or $g$ according to the value of the `coordinate' 1+1. With this structure~\cite{altenkirch2007algebra} defines
\[
\sem{\ite{\hspace{-4pt}\prescript{\circ}{}{\hspace{3pt}p}}{s}{t}} =(\sem{s}\mid \sem{t})\circ (\sem{p}\times \id)\circ \delta_{\Gamma,\Delta}.
\]
We refer the reader to \textit{op.cit.}\hspace{-2pt} for the specific denotational semantics that is based on finite sets; the structural components required to define it are sufficient for our purpose. Note that neither the category $\CPMp$ nor the category $\Hilb$ of finite-dimensional vector spaces possesses this structure (there cannot exist an operator $A\otimes B\to A\times B$), and so the semantics above does not yield the standard semantics of well-known examples of coherent control such as the CNOT or Toffoli gates.
A recent approach axiomatises a notion of coherent control over Props~\cite{heunen2025one} and shows that this corresponds to certain rig categories.
We show in \cref{subsec:vec} that coherent control can also be understood as an instance of \eqref{rule:cond_vanilla} and interpreted accordingly.

\section{\Name categories}\label{sec:def}

In the previous section we have seen four mechanisms for interpreting \ifte statements. None  subsumes the others, they are incomparable. It is therefore natural to ask: what exactly is needed to interpret conditional branching in a general setting?

To answer this question we draw inspiration from the axiomatisation of conditional branching found in the theory of (Guarded) Kleene Algebras with Tests, \ie (G)KATs~\cite{kozen1997kleene,SFHKKS20a}. From this perspective, conditional branching is a collection of binary operations $+_b$ indexed\footnote{We write $p +_b q$ in the style of GKAT~\cite{SFHKKS20a} instead of the more usual KAT expression $bp+\overline{b}q$.} by the elements $b\in B$ of a Boolean algebra of `test programs' and axiomatized by\footnote{\cite{SFHKKS20a} has two more axioms, $e+_b f=be+_b f$ and $eg+_bfg = (e+_b f)g$, from which it is possible to derive the true test axiom $e+_1 f= e$ that we take as fundamental. The first does not apply to our typed setting as $b$ will be a predicate and there is no reason to expect that it can compose with the arbitrary computation $e$. As we shall see the second axiom will always hold when composition is defined.}:

\hspace{-6mm}\begin{minipage}[t]{0.45\textwidth}
	\begin{enumerate}
		\item Skew-commutativity: $e+_b f= f+_{\hspace{1pt}\overline{b}} e$
		\item True test: $e+_1 f= e$
	\end{enumerate}
\end{minipage}\hfill
\begin{minipage}[t]{0.53\textwidth}
	\begin{enumerate}[start=3]
		\item Idempotence: $e+_b e = e$
		\item Skew-associativity: $(e+_b f)+_c g = e+_{bc}(f+_c g)$
	\end{enumerate}
\end{minipage}

\hspace{1pt}

\noindent where $\overline{b}$ stands for the negation of $b$, 1 is the top element of $B$ and $bc$ the conjunction of $b,c\in B$.
Our first task is to adapt these ideas to typed computations with a categorical semantics. In particular, can we generalise the negation operation, the idea of a test returning `true' or even the conjunction operation? As we shall see in \cref{sec:structure}, while we cannot replicate the full boolean structure of deterministic tests in general, the humble coproduct provides us with a lot of the necessary machinery to adapt the axioms above. We therefore assume throughout that our target semantic universe is a \emph{symmetric and distributive monoidal category $(\cat,\otimes, I,+)$} in order to interpret the pattern-matching rule \eqref{rule:pattern} as \eqref{eq:coproduct_semantics}.

Following our terminology for programs, we also call morphisms of the type $X\to I+I$ \emph{predicates} or \emph{tests}. Note that $I$ is in general not the terminal object, in contrast with~\cite{jacobs2015new} which is otherwise similar in spirit to a lot of what follows.
We will call morphisms of the type $X\to X+X$ \emph{branching morphisms} and since the \emph{first} branch of an \ifte statement is the branch corresponding to `true', we will think of the \emph{first} summand of a coproduct $X+X$ as corresponding to `true'. In particular, we will use the convention that the \emph{first} summand of $I+I$ represents the truth value `true', somewhat contrary to convention. Given an object $X$, the following special morphisms $X+X\to X+X$ will be important:
\begin{align*}
	\strue_X \defeq [\inj_1,\inj_1]
	\quad\sfalse_X \defeq [\inj_2,\inj_2]
	\quad\snot_X\defeq [\inj_2, \inj_1]
\end{align*}
We omit subscripts when $X=I$ and write $\strue_X f$ for $\strue_X\circ f$ and similarly for $\sfalse$ and $\snot$.
Objects of the shape $X+X+X+X$ will represent pairs of truth values. The four summands will intuitively correspond to, in order, (true,true), (true, false), (false, true) and (false,false). This justifies the notation for the following special morphisms $X+X+X+X\to X+X$
\begin{align*}
	& \vee_X \defeq [\strue, \id_{1+1}] =  [\inj_1, \inj_1, \inj_1, \inj_2] 
	&
	&\wedge_X \defeq [\id_{1+1}, \sfalse] =  [\inj_1, \inj_2, \inj_2, \inj_2]
\end{align*}
We can now define the categorical structure that will interpret \ifte statements.
\begin{definition}\label{def:name}
	A \textbf{\name category} is a symmetric distributive monoidal category $(\cat, \otimes, I, +)$ equipped with a family of functions
	$
	\cho_X: \cat\big(X, I+I\big) \to \cat\big(X, X+X\big)
	$
	that we will call \emph{choice functions} and which satisfy for every test $p:X\to I+I$
	\begin{enumerate}[nosep]
		\item \emph{Skew-commutativity}: $\cho_X(\snot p) = \snot_X  \cho_X(p)$
		\item \emph{True test}:  $\cho_X(\strue p)=\strue_X\circ \cho_X(p)$
	\end{enumerate}
	A \name structure will be called \emph{idempotent} if it additionally satisfies
	\begin{enumerate}[start=3,nosep]
		\item \emph{Idempotence}: $\nabla_X\circ \cho_X(p)=\id_X$. 
	\end{enumerate}
	Finally, it will be called \emph{commutative} if for every $p, q:X\to I+I$, 
	\begin{enumerate}[start=4,nosep]
		\item\emph{Commutative}: $(\cho_X(q)+\cho_X(q))\circ \cho_X(p) 
		= [\inj_1, \inj_3, \inj_2, \inj_4] \circ (\cho_X(p)+\cho_X(p))\circ \cho_X(q)$.
	\end{enumerate}
\end{definition}
We interpret the \ifte rule \eqref{rule:cond_vanilla}  in any conditionally branching category as
\begin{align}\label{eq:itesemantics}
	\small{
		\xymatrix@C=12ex@R=3ex
		{
			\sem{\Gamma}\otimes \sem{\Delta}
			\ar[r]_-{\cho_{\sem{\Gamma}}(\sem{p})\otimes\id} 
			\ar@/^1.3pc/[rrr]^-{\sem{\Gamma, \Delta\vdash \ite{p}{s}{t}:\type}}
			& 
			(\sem{\Gamma}+\sem{\Gamma})\otimes \sem{\Delta} \ar[r]_{\rdistr}
			& 
			\sem{\Gamma,\Delta}+\sem{\Gamma,\Delta} \ar[r]_-{[\sem{s},\sem{t}]} 
			&
			\sem{\type}
		}
	}
\end{align}
Beyond the simplicity of this semantics, let us provide some justification for our definition. 

\textbf{1. Structural similarity with traces.} Our definition uses the same basic data as that of a traced monoidal category: a monoidal category together with a collection of functions between hom-sets of a particular shape and satisfying certain axioms. This echoes our stated aim to do for \ifte what traced monoidal categories did for recursion and iteration.

\textbf{2. Connection with iteration.} Our definition gives a semantic basis for the interpretation of \texttt{while} loops in the context of Elgot theories and Elgot monads~\cite{elgot1975monadic,adamek2006iterative,adamek2009elgot}. To interpret $x:\type\vdash \mathtt{while}~p~\mathtt{do}~s:\type$ one first needs a morphism $e: \sem{\type}\to \sem{\type}+\sem{\type}$ (possibly effectful) from which to compute the \emph{conditional iterate}~\cite{elgot1975monadic} $e\dg: \sem{\type}\to \sem{\type}$ interpreting the \texttt{while} loop. This is always available in a \name category via $e=(\sem{s}+\id_{\sem{\type}})\circ \cho_{\sem{\type}}(p)$.
The fact that our definition is designed to interpret the \ifte rule \eqref{rule:cond_vanilla}, in which the context can be shared by the guard and the branches, makes it possible to consider iteration even in linear cases (\eg the semantics of probabilistic iteration of~\cite{K79c}). Thus a \name category is always `iteration ready'. This connects to the previous point via the well-know correspondence between iteration and traces for the cocartesian structure~\cite{hasegawa2004uniformity}.

\textbf{3. Connection with instruments.} The shape of our definition has a precedent in quantum control in the shape of the notion of \emph{instrument}, later generalised in \eg \cite{jacobs2015new,jacobs2016affine}. Given a predicate $\Gamma\vdash p: \bit$ defining two orthogonal projections the \emph{L\"{u}ders instrument} \cite{luders2006concerning} provides a splitting operator $\sem{\Gamma}\to\sem{\Gamma}\oplus\sem{\Gamma}$ allowing a semantics of measurement-based control, precisely what a \name structure provides too (see \cref{subsec:vec}).

\textbf{4. GKAT axioms.}
The definition enforces the GKAT axioms (1)-(2) given above. To see this, let us define programs that are syntactic counterparts of morphisms defined above:
\begin{align*}
	&\pnot(p)\hspace{-2pt}\defeq\hspace{-1pt} \caseof{p}\hspace{-2pt}(\pin(1,x)\hspace{-1pt}\Rightarrow \hspace{-1pt}\pin(2,x), \pin(2,x)\hspace{-1pt}\Rightarrow\hspace{-1pt} \pin(1,x))
	\\
	&\ptrue(p)\hspace{-1pt}\defeq\hspace{-1pt} \caseof{p}(\pin(1,x)\hspace{-1pt}\Rightarrow \hspace{-1pt}\pin(1,x), \pin(2,x)\hspace{-1pt}\Rightarrow\hspace{-1pt} \pin(1,x))
\end{align*}
With these programs we can state and show:
\begin{theorem}\label{thm:semantic_identites}
	The interpretation \eqref{eq:itesemantics} of \ifte statements in a conditionally branching category guarantees the following semantic identities:
	\begin{enumerate}[nosep]
		\item $\sem{\Gamma\hspace{-1pt}\vdash\hspace{-1pt} \ite{p}{s}{t}:\type} \hspace{-1pt}=\hspace{-1pt}\sem{\Gamma\hspace{-1pt}\vdash\hspace{-1pt} \ite{\hspace{-1pt}\pnot p}{t}{s}:\hspace{-1pt}\type}$
		\item $\sem{ \Gamma\vdash \ite{\ptrue(p)}{s}{t}:\type} = \sem{\Gamma\vdash s:\type}$.
	\end{enumerate}
	If the \name structure is idempotent, then
	\begin{enumerate}[start=3, nosep]
		\item $\sem{\Gamma\vdash \ite{p}{s}{s}:\type}=\sem{\Gamma \vdash s:\type}$.
	\end{enumerate}
	If the \name structure is commutative then
	\begin{enumerate}[start=4, nosep]
			\item $\sem{\Gamma\hspace{-1pt}\vdash\hspace{-1pt} \ite{p}{\ite{q}{s_1}{s_2}}{\ite{q}{s_3}{s_4}}:\hspace{-1pt}\type}\hspace{-1pt}= \\ \sem{\Gamma\vdash \ite{q}{\ite{p}{s_1}{s_3}}{\ite{p}{s_2}{s_4}}:\type}$.
	\end{enumerate}
\end{theorem}
The reader may wonder why skew-associativity has been left out of \cref{thm:semantic_identites}. The reason is that the obvious generalisation of skew-associativity does not hold in general. We will look at nested conditional branchings and skew-associativity in detail in \cref{sec:skew}, but we can give a simple example to illustrate the problem. For purely probabilistic tests like $\progfont{flip}~\nicefrac{1}{2}$, it is in general impossible to find an $r\in [0,1]$ playing the role on the conjunction of guards in GKAT's skew-associativity and making the following skew-associativity identity hold:
\begin{align*}
	&
	\sem{\ite{\progfont{flip}~q}{(\ite{\progfont{flip}~p}{s_1}{s_2})}{s_3}} 
	\\
	=~ & \sem{\ite{\progfont{flip}~r}{s_1}{(\ite{\progfont{flip}~q}{s_2}{s_3}})}
\end{align*}
The reason is that, semantically, this would require the probability of reaching $s_3$ to be $(1-q)=(1-q)(1-r)$ which does not hold in general. This suggests that the skew-associativity rule from GKAT, if it is to hold in more generality, must be a special case of a more general rule $(e+_b f)+_c g = e+_{bc}(f+_{b\bullet c} g)$ for some operation $b\bullet c$ yet to be determined and evaluating to $c$ in the case of deterministic boolean tests. We return to this question in \cref{sec:skew}.
We can however show the identity corresponding to the GKAT axiom $eg+_b fg=(e+_bf)g$. 
For this we will use the standard \texttt{let}-binding rule
\[
\small{
\infer{\Gamma,\Delta\vdash \progfont{let}~x=s~\progfont{in}~ t:\type}{\Gamma\vdash s:\type[S] & \Delta, x:\type[S]\vdash t:\type}
}
\]


\begin{proposition}\label{prop:distributivity}
	In any \name category, the interpretation \eqref{eq:itesemantics} satisfies
	{\small
		\begin{align}
			& \sem{\Gamma,\Delta\vdash \ite{p}{\progfont{let}~x=s_1~\progfont{in}~t}{\progfont{let}~x=s_2~\progfont{in}~t}:\type}\nonumber
			\\
			=~&\sem{\Gamma,\Delta\vdash \progfont{let}~x=(\ite{p}{s_1}{s_2})~\progfont{in}~t:\type}\label{eq:GKATdistributivity}
		\end{align}
	}			
	(where $\mathtt{let}$ bindings are interpreted in the usual way).
\end{proposition}

\begin{remark}
	It would seem natural to require that choice functions satisfy some naturality condition (no pun intended). The only one that makes sense is to ask that for all $f: Y\to X$, 
	\begin{align}\label{eq:naturality}
		\cho_X(p)\circ f = (f+f)\circ \cho_Y(p\circ f).
	\end{align}
	Although this condition does hold for many \name categories, it does not hold in some important situations, for example non-deterministic computations. 
	Unlike the definition of a trace, the \name structure doesn't seem to demand any naturality conditions. However, the skew-commutativity and true test axioms do imply a kind of `combinatorial naturality' in $I+I$ in the sense that every `combinatorial' endomorphism, that is to say the 4 endomorphisms $f_I: I+I\to I+I$ generated by the coproduct structure, must be reflected on $X+X$ via $\cho_X(f_I\circ p) = f_X\circ \cho_X(p)$. 
\end{remark}

\begin{remark}
	It follows from \cref{def:name} that if a category $\cat$ is equipped with a conditional branching structure, so is every subcategory $\cat[D]$ of $\cat$ containing $I$ and closed under $\otimes$ and +.
\end{remark}

We conclude this section by considering the notion of \emph{short-circuiting} (\eg~\cite[p.222]{pierce2002types}). Short-circuiting is a behaviour of conditional branching characterised by the fact that classically \emph{only one of the branches ever needs to be evaluated}: the if-branch when the guard evaluates to \texttt{true}, and the else-branch otherwise. In both cases, the remaining branch is `short-circuited' \ie not evaluated. 

\begin{definition}\label{def:shortcircuit}
	A \name structure on a category $\cat$ will be called \emph{short-circuiting} if for every state $s: I\to X$  and every test $p: X\to I+I$, if $p\circ s$ factors through $\inj_1: I\to I+I$, then the morphism $\cho_{X}(p)\circ s$ factors through $\inj_1: X\to X+X$, and similarly for $\inj_2$. Diagrammatically:
	\begin{align*}
		\text{if} \quad
		\xymatrix@C=12ex@R=3ex{
			I\ar[r]^{s}\ar@{..>}[d] & X\ar[d]^{p}
			\\
			I\ar[r]_{\inj_1\text{ or }\inj_2} & I+I
		}
		\quad
		\text{then}
		\quad
		\xymatrix@C=12ex@R=3ex{
			I\ar@{..>}[d] \ar[r]^{s} & X\ar[d]^{\cho_{X}(p)} 
			\\
			X\ar[r]_{\inj_1\text{ or }\inj_2} & X+X
		}
	\end{align*}
\end{definition}
The definition guarantees that whenever a predicate factors through $\strue$ in a state $s$, then the semantics of the \ifte statement can be written by only evaluating the semantics of the if-branch (since $\big[\sem{s},\sem{t}\big]\circ \inj_1=\sem{s}$), and similarly for the else-branch.

\section{Examples of \name categories}\label{sec:examples}


\subsection{Distributive copy-categories}\label{subsec:namecopy}
We first present a large class of categories in which the copy-and-distribute strategy discussed in \cref{subsec:classical} can be used to define a \name structure. We do not need (Cartesian) distributive categories -- as we saw in \cref{subsec:classical} this excludes semantically important categories -- but we do need categories that are distributive monoidal as a prerequisite to  \cref{def:name}.

\begin{example}
	We saw in \cref{ex:Rel} that the category $\Rel$ of non-deterministic computations is not distributive \wrt the Cartesian product of $\Rel$. It is however distributive \wrt the monoidal structure given by the Cartesian product in $\Set$ with $I=\{\ast\}$ the singleton set. 
\end{example}

\begin{example}
	Similarly, we showed in \cref{ex:CPO} that the category $\oCPOs$ of pointed $\omega$-cpos and strict morphisms is not distributive \wrt the Cartesian product. It is however distributive \wrt the \emph{smash product} with carrier $A\otimes B=(A\setminus \{\bot_A\})\times (B\setminus\{\bot_B\})\cup \{\bot\}$ and the pointwise order. The unit of this product is the chain $\{\bot\leq \ast\}$.
\end{example}

In order to apply the copy-and-distribute method of \cref{subsec:classical} we also need to be able to copy (a functionality that comes for free with the Cartesian product).  The two examples above are examples of distributive \emph{copy categories} \cite{cockett2002restriction}, \ie symmetric monoidal categories where every object comes equipped with a cocommutative and coassociative copying morphism $\cp_A: A\to A\otimes A$. 
In $\Rel$ the copy map is given by the diagonal map $\cp_A(x)= (x,x)$, in $\oCPOs$ it is given by $\cp_A(x) = (x,x)$ if $x\neq \bot$ and $\bot$ else.
\begin{theorem}\label{thm:copydistr}
	Every distributive copy category has a commutative conditionally branching structure given by the choice functions
	\begin{align}\label{eq:copydistr}
		\cho_{X}:~\cat(X, I+I)&\to\cat(X, X+X),\nonumber \\
		p&\mapsto  (\rho_X+\rho_X)\circ \ldistr_X\circ (\id\otimes p)\circ \cp_X: \nonumber 
		\\
		X&\to X\otimes X\to X\otimes (I+I)\to X\otimes I+X\otimes I\to X+X
	\end{align}
	This structure is short-circuiting in all deterministic states $s: I\to X$ (\ie states such that $\cp_X\circ s=(s\otimes s)\circ \cp_I $, see~\cite{fritz2020synthetic}). 
\end{theorem}
The conditionally branching structure of \eqref{eq:copydistr} is in general \emph{not} idempotent, therefore it does not support the rewriting schema 
$\sem{\Gamma\vdash \ite{p}{s}{s}:\type}=\sem{\Gamma \vdash s:\type}$. To recover this behaviour, we need more structure on $\cat$. A \emph{copy-discard category}~\cite{cho2019disintegration} is a copy category such that every object also comes equipped with a deletion morphism $\del_A: A\to I$ which, together with the copy morphism $\cp_A: A\to A\otimes A$, equips it with a comonoid structure. If the unit $I$ is terminal, then a copy-delete category is called a Markov category~\cite{fritz2020synthetic}.

\begin{theorem}\label{thm:markov}
	The conditionally branching structure given by \eqref{eq:copydistr} is idempotent whenever the distributive copy category $\cat$ is Markov, in particular when it is distributive (\cref{subsec:classical}).
\end{theorem}

\begin{example}
	The category $\Stoch$ \cite[\S 4]{fritz2020synthetic} of measurable spaces and Markov kernels is a Markov category, and can therefore be equipped with an idempotent and commutative \name structure following \eqref{eq:copydistr}. It sends a stochastic test $p: X\to \mathcal{G}(1+1)$, where $\mathcal{G}$ is the Giry monad, to $\cho_{X}(p): X\to \mathcal{G}(X+X)$ given by
	\[
	\cho_{X}(p)(x)=p(x)(1)\delta_{\inj_1 x}+p(x)(2)\delta_{\inj_2 x}
	\]
\end{example}

\begin{example}
	The category $\Rel$ is a copy category but it is not a Markov category as the terminal object (which is also the initial object by self-duality) is the empty set which is not the unit of the monoidal product described above. $\Rel$ therefore supports a commutative conditional branching structure, but not an idempotent one. This comes from the fact that the empty relation $\emptyset: X\to I+I$ semantically represents a crashing/divergent test, and for such a test it cannot be the case that $\sem{\Gamma\vdash \ite{p}{s}{s}:\type}=\sem{\Gamma \vdash s:\type}$.
\end{example}

\begin{example}
	Similarly, the category $\oCPOs$ is a copy category but not a Markov category as the unit of the smash product is not terminal. The commutative conditional branching structure of $\oCPOs$ given by \eqref{eq:copydistr} cannot be idempotent for the same reason as in the previous example: predicates $p: X\to I+I$ allow for divergence (that is the point of the bottom element $\bot$) and it therefore cannot be the case that $\sem{\Gamma\vdash \ite{p}{s}{s}:\type}=\sem{\Gamma \vdash s:\type}$.
\end{example}

\begin{remark}
	The example above illustrates the fact that for a call-by-value semantics allowing divergence, \eg in the category $\oCPOs$, a conditionally branching structure can never be idempotent, and $\sem{\Gamma\vdash \ite{p}{s}{s}:\type}=\sem{\Gamma \vdash s:\type}$ will not hold. It is however also worth noting that even in a call-by-name semantics (where $f\bot=\bot$ needn't hold), guards behave in a call-by-value way. Specifically, the call-by-name small-step  operational semantics of an \ifte statement forces the evaluation of the guard and only starts evaluating one of the branches if the guard evaluates to \texttt{true} or \texttt{false} (in particular it is short-circuiting). This means that if the guard diverges, so does the entire \ifte statement, in contrast with $f\bot\neq \bot$ in general. The idempotence of a \name structure is thus prevented by partiality, no matter what the evaluation strategy is. 
\end{remark}


\begin{proposition}\label{prop:cdcat_shortcrcuit}
	The \name structure of \eqref{eq:copydistr} is short-circuiting in any copy-discard category $\cat$ where the following diagram is a pullback:
	\[
	\xymatrix@C=10ex@R=3ex{
		& X\ar[r]_{\inj_1} \ar[d]_{\del}
		& X+X\ar[d]^{\del+\del}
		\\
		& I\ar[r]^{\inj_1}
		& I+I
	}
	\]
\end{proposition}
This result does not require the copy-discard category to be Markov. In the Cartesian case it is a special case of~\cite[(4.4)]{jacobs2015new} and therefore holds more generally in any effectus~\cite[\S 4]{jacobs2015new}. 
\newcommand{\str}{\mathrm{str}}
\textbf{Effectful Computations} interpreted in the Kleisli categories $\cat_T$ of monads $T:\cat\to\cat$ provide a rich collection of examples of \name structures given by \eqref{eq:copydistr}. 
Let us start by examining when $\cat_T$ is distributive monoidal -- the prerequisite for \cref{def:name} -- assuming that $\cat$ is. First, $\cat_T$ always has coproducts when the base category $\cat$ does, with inclusions given by $\eta^T\circ\inj_1, \eta^T\circ\inj_2$ and copairings $A+B\to TC$ built in the base category. 
To define a monoidal structure on $\cat_T$ we start by recalling that a \emph{left strength} for a monad $T$ on $(\cat, \otimes, I)$ is a natural transformation $\str_{A,B}: A\otimes TB\to T(A\otimes B)$ satisfying axioms found in \eg \cite{kock1972strong,moggi1991notions}. If $\cat$ has a symmetry transformation $\gamma_{A,B}: A\otimes B\to B\otimes A$, we can define the \emph{right strength}
$\str'_{A,B}=T(\gamma_{B,A})\circ \str_{B,A}\circ \gamma_{TA,B}: TA\otimes B\to T(A\otimes B)$.
Finally, a monad with a left strength is \emph{commutative} whenever the following diagram commutes:
\[
\xymatrix@C=10ex@R=3ex{
	TA\otimes TB \ar[r]_-{\str_{TA,B}}\ar[d]_{\str'_{A,TB}}
	&
	T(TA\otimes B)\ar[r]_-{T\str'_{A,B}}
	&
	T^2(A\otimes B)\ar[d]^{\mu_{A\otimes B}}
	\\
	T(A\otimes TB) \ar[r]^{T\str_{A,B}}
	&
	T^2(A\otimes B)\ar[r]^{\mu_{A\otimes B}}
	&
	T(A\otimes B)
}
\] 
The transformation $TA\otimes TB\to T(A\otimes B)$ makes $T$ a monoidal monad (there is a one-to-one correspondence between commutative monads and symmetric monoidal monads~\cite{kock1972strong,dahlqvist2018layer}). It follows that
the Kleisli category of a commutative monad on a symmetric monoidal category is symmetric monoidal in a canonical way \cite[Cor.\ 4.3]{power1997premonoidal}. Finally, it is not hard to check that:
\vspace{-5mm}
\begin{proposition}\label{prop:kleisli_distrib}
	If $(\cat, \otimes, I, +)$ is a symmetric distributive monoidal category and $T: \cat\to\cat$ is a commutative monad, then $\cat_T$ is symmetric distributive monoidal with the distributivity maps given by the $\cat$-morphisms $\eta^T\circ \ldistr$ and  $\eta^T\circ \rdistr$.
\end{proposition}
With this in place, we can apply the copy-and-distribute approach to defining \name structures on Kleisli categories. Assuming that $(\cat, \otimes, I, \cp)$ is a distributive copy category and that $T$ is commutative, we can define copy morphisms on $\cat_T$ by post-composition with the unit: $\eta^T_X\circ \cp_X: X\to T(X\otimes X)$. This gives us a commutative \name structure on $\cat_T$ via \eqref{eq:copydistr} and \cref{thm:copydistr}. After some simplifications the choice function associated with a $\cat_T$-predicate $p: X\to I+I$ is explicitly given by
\vspace{-1mm}
\begin{align}\label{eq:liftedcopydistr}
	\cho_{X}(p) =& T\ldistr_X\circ  \str_{X, I+I}\circ (\id_X\otimes p)\circ \cp_X: \\
	& X\to X\otimes X\to X\otimes T(I+I)\to T(X\otimes (I+I))\to T(X+X)\nonumber
\end{align}
\vspace{-1.3\baselineskip}
%

The question of the \name structure's idempotence is once again more subtle and we require quite a bit more structure on the base category and on the monad to establish it. This question is discussed in some detail in~\cite[\S5]{jacobs2016affine} where idempotence is called \emph{side-effect-freeness} and the monoidal structure is given by the Cartesian product and the terminal object. We generalise this discussion to the monoidal case. Recall that a monad on a category $\cat$ with a terminal object $1$ is called \emph{affine} if $T1\simeq 1$. 
\vspace{-1mm}
\begin{theorem}[\cite{fritz2020synthetic}]
	Let $\cat $ be a Markov category, and let $T$ be a commutative monad on $\cat$ that is also affine. Then the Kleisli category $\cat_T$ is also a Markov category where deleting is given by $\del_A^T=\del_A: A\to 1\simeq T1$.
\end{theorem}
\vspace{-4mm}
\begin{corollary}\label{cor:monad_idempotent}
	Let $\cat $ be a Markov category, and let $T$ be a symmetric monoidal affine monad on $\cat$, then the \name structure defined in \eqref{eq:liftedcopydistr} is idempotent.
\end{corollary}
\vspace{-2mm}
Most useful monads are not affine, and therefore do not support an idempotent \name structure. We can however prove idempotence for pure tests (tests factoring through the unit $\eta^T$), whenever the base category is idempotent \name.
\vspace{-1mm}
\begin{theorem}\label{thm:pure_tests}
	Let $\cat$ be an idempotent \name category, and $T:\cat\to\cat$ be a monad with unit $\eta^T$. For pure tests $\eta^T\circ p:\hspace{-1pt} X\hspace{-1pt}\to\hspace{-1pt} I+I\hspace{-1pt}\to\hspace{-1pt} T(I+I)$, the branching morphisms
	\vspace{-1mm}
	\begin{align*}
	\widehat{\cho}_{X}(p) = \eta^T_{X+X} \circ \cho_{X}(p): X \to X+X\to T(X+X)
	\end{align*}
	satisfy axioms (1)-(3) of \cref{def:name}.
\end{theorem}
\vspace{-2mm}
\subsection{Categories of Vector Spaces}\label{subsec:vec} 
In this section we present a second class of categories with a \name structure, this time based on the ability to define choice functions as `splitting morphisms'.

\textbf{Probabilistic Program Semantics.}  We start with our motivating example and prove the existence of a \name structure for a large class of real Banach spaces which includes the denotation of types in~\cite{K79c}. The intuition behind the construction is that a predicate $p:X\to \R\oplus\R$ defines a linear functional $ev_p: \nabla_{\R}\circ p: X\to \R$. Using a generalisation of the Radon-Nikodym theorem, one can compute the `densities' of the true and false components $\pi_1\circ p, \pi_2\circ p$ \wrt $ev_p$, and use these to split any positive input vector.
\vspace{-5mm}
\begin{theorem}\label{thm:vector_split}
	Let $X$ be a $\sigma$-order complete ordered Banach space 
	and let $p: X\to \R\oplus\R$ be a norm-preserving $\sigma$-order continuous positive operator. For any $0\leq x\in X$ there exist $0\leq x_t, x_f\in X$ s.th.
	\begin{enumerate}[nosep]
		\item $x=x_t+x_f$
		\item $x\mapsto (x_t, x_f)$ defines a positive linear operator $\cho_X: X\to X\oplus X$
		\item $\cho_X$ satisfies the axioms (1)-(4) of \cref{def:name} and is short-circuiting
		\item If $X$ is an AL-space, then $\norm{x_t}=\pi_1\circ p(x)$ and $\norm{x_f}=\pi_2\circ p(x)$
	\end{enumerate}
\end{theorem}This result generalises the semantics of~\cite{K79c} to arbitrary probabilistic guards (and corrects an erroneous splitting in \cite{dahlqvist2019semantics}). For a deterministic guard $f_\ast: \Meas(X_1\times \ldots \times X_n)\to \Meas(1+1)$ and an input distribution $\mu$ we recover the split $\mu\mapsto (\one_{f\inv(0)}\cdot \mu, \one_{f\inv(1)}\cdot \mu)$, where $f\cdot\mu$ is the measure with density $f$ \wrt to $\mu$. For an arbitrary guard we get the split $\mu\mapsto (f_{\mathrm{if}}\cdot \mu, f_{\mathrm{else}}\cdot \mu)$ where $f_{\mathrm{if}}(x_1, \ldots, x_n)=p(\delta_{x_1}\otimes \ldots\otimes \delta_{x_n})(1)$, the probability of `true' at $(x_1, \ldots,x_n)$, and similarly for $f_{\mathrm{else}}$. Thus $\sem{\mathtt{flip}~\nicefrac{1}{2}}:\Meas(X_1\times \ldots \times X_n)\to \Meas(1+1)$ yields $\mu\mapsto (\nicefrac{\mu}{2},\nicefrac{\mu}{2})$.

\vspace{1mm}
\textbf{Measurement-based quantum control} can be understood as an instance of our main \ifte rule \eqref{rule:cond_vanilla} interpreted by a \name structure on the category $\CPMp$ provided by the notion of L\"{u}ders instrument~\cite{luders2006concerning}. Given a predicate $p: X\to \C\oplus \C$, we can define two operators $X\to X$ via $I_p = p\dg \pi_1 p$ and $E_p=p\dg \pi_2 p$
where $p\dg$ is the adjoint of $p$ and $\pi_i:\C\oplus\C\to\C\oplus\C$ is the projection onto the $i^{th}$ component. The notion of L\"{u}ders instrument then provides us the splitting map $\cho_{X}(p):X\to X\oplus X, \rho \mapsto (\sqrt{I_p}\rho \sqrt{I_p}, \sqrt{E_p}\rho \sqrt{E_p})$. With this, we can revisit \cref{ex:meas_based_ifte} and type the program as
\[
\small{
	\infer{x,y:\qubit\vdash \ite{\meas(x)}{x\otimes y}{x\otimes \progfont{Z}y}:\qubit\otimes \qubit}
	{x:\qubit\vdash\meas(x):\bit & x,y:\qubit\vdash x\otimes y:\qubit\otimes \qubit & x,y:\qubit\vdash x\otimes \progfont{Z}y:\qubit\otimes \qubit}
}
\]
The instrument corresponding to $\meas(x)$ is given by the projections $I_{\meas(x)} = \ket{0}\bra{0}$ and $E_{\meas(x)}=\ket{1}\bra{1}$ yielding the semantics $\sem{\ite{\meas(x)}{x\otimes y}{x\otimes \progfont{Z}y}}$ given by \eqref{eq:itesemantics}:
{\small
\begin{align*}
	\hspace{-4mm}
	\begin{pmatrix}A & B \\ C& D\end{pmatrix}\mapsto & \big[ \id\otimes \id, \id\otimes \sem{\progfont{Z}}\big]\circ (\cho(\sem{\meas})\otimes \id) \begin{pmatrix}A & B \\ C& D\end{pmatrix}
	= \begin{pmatrix}A & 0 \\ 0& ZDZ\dg\end{pmatrix}
\end{align*}
}The key difference with the semantics of \cref{subsec:linear} is that the qubit $x$ remains accessible (in its collapsed state) to the computation. The guard modifies its context, it does not consume it.\begin{remark}
	The construction above relies on the fact that Hilbert spaces are self-dual. With some care one can build adjoint pairs between $\mathrm{L}^1(X,\mu)$ and $\mathrm{L}^\infty(X,\mu)$ for a measured space $(X,\mu)$~\cite{dahlqvist2018borel}. The construction above can then be applied to probabilistic semantics involving measured spaces, \eg spaces with a prior~\cite{dahlqvist2019semantics}. In that case, $I_p$ would \emph{condition} on observing `true' (\wrt a prior $\mu$), relating conditional branching with probabilistic conditioning.
\end{remark}

\textbf{Coherent quantum control} can also be understood as an instance of rule \eqref{rule:cond_vanilla} interpreted via a \name structure, but unfortunately not in the same category as above ($\CPMp$ is in some sense designed to tackle measurements). Consider the category $\Hilb$ of finite-dimensional Hilbert spaces and bounded operators (we cannot restrict to contractions as coproducts would then not exist). We interpret coherent control in this category in much the same way as we did above: given $p: X\to \C\oplus\C$, define $I_p, E_p$ as above and the choice functions $\cho_X(p): X\to X\oplus X$ as $x\mapsto (I_p x, E_p x)$. With this `instrument' we can interpret the CNOT gate represented as the program
\[
\small{
	\infer{x,y:\qubit\vdash \ite{x}{x\otimes y}{x\otimes \pnot y}:\qubit\otimes \qubit}
	{x:\qubit\vdash x:\bit & x,y:\qubit\vdash x\otimes y:\qubit\otimes \qubit & x,y:\qubit\vdash x\otimes \pnot y:\qubit\otimes \qubit}
}
\]
where we tacitly use the trivial/identity test $x:\qubit\vdash x:\bit$ which exploits the fact that $\sem{\qubit}=\sem{\bit}=\C^2$ in $\Hilb$. The semantics is then given by $\big[ \id\otimes \id, \id\otimes \sem{\pnot}\big]\circ \langle I_p,E_p\rangle$ that splits
$a_{00}\ket{00}+a_{01}\ket{01}+a_{10}\ket{10}+a_{11}\ket{11}$ into $(a_{00}\ket{00}+a_{01}\ket{01}, a_{10}\ket{10}+a_{11}\ket{11})$ and then applies both branches in superposition to yield $a_{00}\ket{00}+a_{01}\ket{01}+a_{10}\ket{11}+a_{11}\ket{10}$  as expected. Note that unlike \eqref{rule:QML_coherent}, the context of the guard is shared with the branches.

\begin{proposition}\label{prop:quantum_split}
	The choice functions defined above define a commutative and idempotent  \name structure on $\Hilb$ for \emph{sharp} predicates, \ie predicates satisfying $\mathrm{im}(I_p)=\ker(E_p)$.
\end{proposition}

\begin{remark}
	Unifying the two types of control semantically is notoriously difficult~\cite{valiron2022semantics}. $\Hilb$ cannot deal with measurements and, conversely, applying the L\"{u}ders instrument of $\CPMp$  to coherent control yields the wrong semantics, \eg in the CNOT case it would ignore the cross-terms $B,C$. We settle for the structural unification provided by \eqref{rule:cond_vanilla} and \eqref{eq:itesemantics}.
\end{remark}

\section{Structure of branching morphisms and tests}\label{sec:structure}
Having defined \name categories as a simple categorical structure in which to interpret \ifte statements via `branching morphisms', we now turn our attention to studying the algebraic structure of branching morphisms and of tests in a \name category. 

\subsection{Structure of the sets $\cat(X, X+X)$ of branching morphisms}\label{subsec:structure_bm}

In \cref{sec:def} we defined three special morphisms: $\strue_X, \sfalse_X, \snot_X:X+X\to X+X$ which can be viewed as defining unary operations on $\cat(X,X+X)$ by post-composition. We can also define two constants on $\cat(X,X+X)$: $\top_X=\inj_1: X\to X+X$ and $\bot_X=\inj_2: X\to X+X$.
What about binary operations on branching morphisms? Such operations seem essential if we are to understand nested \ifte statements branching from $X$ to $X+X+X+X$.

If we view branching morphisms $X\to X+X$ as morphisms $X\to TX$ for the functor $T=\Id+\Id$ and manage to equip $T$ with a monad structure, then this would also equip the set $\cat(X,X+X)$ -- viewed as the set $\Kl{T}\!(X,X)$ --  with an operation for combining nested branchings. To see which monad structures on $T=\Id+\Id$ are useful, we list all such structures in the case of the category $\Set$.


\begin{theorem}\label{thm:monad_struct_set}
	The following unit-multiplication pairs define monad structures on the functor $T=\Id+\Id$ in any category with coproducts. 
	\begin{multicols}{2}
		\begin{enumerate}[nosep]
			\item $\eta = \inj_1, \mu=[\id,\snot]$
			\item $\eta = \inj_1, \mu=\wedge$
			\item $\eta = \inj_2, \mu=[\snot, \id]$
			\item $\eta = \inj_2, \mu=\vee$
		\end{enumerate}
	\end{multicols}
	\noindent In $\Set$ these are the \emph{only} possible monad structures on $T$.
\end{theorem}
It follows that $\cat(X,X+X)$ can always be equipped with four monoidal structures. For $f,g\in \cat(X,X+X)$ we write $f\wedge g$ for the branching morphism $\wedge \circ (g+g)\circ f$ (we drop subscripts in the interest of clarity), and similarly for $\vee$.
For branching morphisms generated by tests in a \name category $\vee,\wedge$ have the following algebraic structure.\begin{proposition}\label{prop:kleisli_structure_tests}
	For any tests $p,q,r: X\to I+I$ in a \name category,
	\begin{multicols}{2}
	\begin{enumerate}[nosep]
		\item $\cho(p)\vee(\cho(q)\vee \cho(r))=(\cho(p)\vee(\cho(q))\vee \cho(r)$
		\item $\cho(p)\wedge(\cho(q)\wedge \cho(r))=(\cho(p)\wedge(\cho(q))\wedge \cho(r)$
		\item $\snot(\snot \cho(p))=\cho(p)$
		\item $\snot(\cho(p)\wedge\cho(q))=\cho(\snot p)\vee \cho(\snot q)$
		\item $\snot(\cho(p)\vee\cho(q))=\cho(\snot p)\wedge \cho(\snot q)$
	\end{enumerate}
	\end{multicols}
	\noindent If the \name structure is idempotent then 
	\begin{multicols}{2}
		\begin{enumerate}[start=6]
			\item $\top\vee \cho(p)=\cho(p)\vee\top=\top$
			\item $\bot\wedge \cho(p)=\cho(p)\wedge\bot=\bot$
			\end{enumerate}
	\end{multicols}
	\noindent If the \name structure is commutative then 
	\begin{multicols}{2}
	\begin{enumerate}[start=8]
		\item $\cho(p)\vee \cho(q) = \cho(q)\vee \cho(p)$
		\item $\cho(p)\wedge \cho(q) = \cho(q)\wedge \cho(p)$
	\end{enumerate}
	\end{multicols}
\end{proposition}

\begin{remark}
	Since $\snot$ is involutive and relates $\vee$ and $\wedge$ via de Morgan laws, $\wedge$ can be defined in terms of $\vee$ and vice-versa. The basic structure (items 1-5) of branching morphisms is thus that of an involutive magma.
	It is tempting to assume that the operators $\vee$ and $\wedge$ satisfy other boolean identities. Probabilistic computations provides some easy counter-examples. Let $p,q,r: X\to \mathcal{D}(1+1)$ be probabilistic tests and let us write $p$ for $p(x)(1)$.
	\begin{enumerate}
		\item $\cho(p)\wedge \cho(p) = \cho(p)$ doesn't hold since it would imply $pp=p$ which isn't generally true. 
		\item Similarly, $\cho(p)\vee \cho(p) = \cho(p)$ doesn't hold since it would imply $(1-p)(1-p)=(1-p)$. 
		\item $\cho(p)\vee \cho(\snot p) = \top$ and $\cho(p)\wedge \cho(\snot q) = \bot$ do not hold since $p(1-p)\neq 0$ in general.
		\item $\cho(p)\wedge (\cho(q)\vee \cho(r))=(\cho(p)\wedge \cho(q))\vee (\cho(p)\wedge \cho(r))$ does not hold as it would imply that 
		\[
		p(1-(1-q)(1-r)) = pqpr + pr(1-pr)+(1-pq)pr
		\]
		which does not hold in general. The other distributivity laws similarly does not hold.
	\end{enumerate}
\end{remark}


\subsection{Structure of the sets $\cat(X, I+I)$ of predicates}\label{subsec:structure_tests}

We can use the \name structure on a category $\cat$ to define binary operations on the set of tests $\cat(X,I+I)$ as follows. For $p,q: X\to I+I$, we overload $\vee, \wedge$ and define:

\vspace{-3mm}
\begin{minipage}{0.43\textwidth}
	\begin{equation}
		\begin{aligned}
			\hspace{-6mm} p\vee q \defeq \vee \circ (q+q)\circ \cho_X(p) \label{def:join_test}
		\end{aligned}
	\end{equation}
\end{minipage}
\begin{minipage}{0.50\textwidth}
	\begin{equation}
		\begin{aligned}
			\qquad p \wedge q \defeq \wedge \circ (q+q)\circ \cho_X(p)\label{def:meet_test}
		\end{aligned}
	\end{equation}
\end{minipage}

\hspace{1pt}

\noindent With these operations we recover part of a familiar algebraic structure on sets of predicates.

\begin{proposition}\label{prop:commutative}
	Predicates on an object $X$ in a \name category form an involutive magma $(\cat(X, I+I), \wedge, \snot)$ (\resp $(\cat(X, I+I), \vee, \snot)$).
	If the \name structure is commutative, so is $\wedge$ (\resp $\vee$).
\end{proposition}

\begin{example}[The boolean algebra of non-deterministic tests]\label{ex:nondet_test}
	Consider finitely branching non-deterministic computations given by the finite powerset monad $\mathcal{P}_f$ on $\Set$. The corresponding choice functions are given by \eqref{eq:liftedcopydistr}. It is straightforward to compute \eqref{def:join_test}-\eqref{def:meet_test} and we get for $p,q: X\to \mathcal{P}_f(1+1)$ that $1\in (p\vee q)(x)\Leftrightarrow 1\in p(x)\cup q(x)$ and $1\in (p\wedge q)(x)\Leftrightarrow 1\in p(x)\cap q(x)$, and dually for 2 (\ie false) $2\in (p\vee q)(x)\Leftrightarrow 2\in p(x)\cap q(x)$ and $2\in (p\wedge q)(x)\Leftrightarrow 2\in p(x)\cup q(x)$.
\end{example}

\begin{example}[The involutive magma of probabilistic tests]\label{ex:prob_test}
	Consider now finitely branching probabilistic computations given by the distribution monad $\mathcal{D}$ on $\Set$, again with the choice functions defined by \eqref{eq:liftedcopydistr}. Given probabilistic tests $p,q: X\to \mathcal{D}(1+1)$, we compute
	\begin{align*}
		(p\wedge q)(x)(1) = p(x)(1)q(x)(1) \qquad
		(p\vee q)(x)(1) = p(x)(1) + q(x)(1) - p(x)(1)q(x)(1)
	\end{align*}
	Thus probabilistic tests carry the standard probabilistic structure on $[0,1]$ (sometimes know as product $t$-norm and sum $t$-conorm in fuzzy logic, pointwise, when evaluated at 1, and the dual structure when evaluated at 2.
\end{example}


By \cref{def:name}, choice functions $\cho_X: \cat(X,I+I)\to\cat(X, X+X)$ always commute with the unary operators $\strue, \snot$.  
In some instance we can also prove that they act as monoid homomorphisms between the $\vee$- and $\wedge$-structures described in \cref{prop:kleisli_structure_tests,prop:commutative}. 
\begin{proposition}\label{prop:homomorphism}
	If $\cho_X(p\vee q)=\cho_X(p)\vee \cho_X(q)$ and $\cho_X(p\wedge q)=\cho_X(p)\wedge \cho_X(q)$ then both $(\cat(X, I+I), \wedge, \snot)$ and $(\cat(X, I+I), \vee, \snot)$ form involutive magmas. If the \name structure is commutative, they are also commutative.
	This holds for the \name structures defined by \eqref{eq:copydistr}.
\end{proposition}
Not all the structure of $\cat(X,X+X)$ can carry over to $\cat(X,I+I)$. The constant branching morphisms $\top,\bot$ have in general no test equivalents since there is in general no distinguished morphism $X\to I$; some objects in $\Ban$ for example have no such morphisms.
\subsection{Skew-associativity}\label{sec:skew}
When the structure on the set of tests is sufficiently rich, we can generalise the skew-associativity rule of GKAT to more general types of computations. The extra structure needed is the following. 
We define the relation $\lesssim$ on $(\cat(X,I+I), \wedge)$ by $p\lesssim q\Leftrightarrow \exists r: p=q\wedge  r$. The relation is transitive whenever  $\wedge$ is associative, \ie whenever $(\cat(X,I+I), \wedge)$ is a magma. 
We will say that $(\cat(X,I+I), \wedge)$ is \emph{conditionally divisible} if for all tests $p,q\in \cat(X,I+I)$ such that $p\leq q$ there exists a test $\frac{p}{q}\in \cat(X,I+I)$ such that $q\wedge\frac{p}{q}=p$. 

Interestingly, we will need the same technical condition on the base category $\cat$ as made in~\cite[Eq. (4.3)]{jacobs2015new}, namely that the maps $
[\id_X,\nabla_X], [\snot_X, \id_X]: X+X+X\to X+X$
are jointly monic. This condition is satisfied in $\Set$ and therefore in any concrete category.
\begin{theorem}\label{thm:skew-assoc}
	Let $\cat$ be a \name category satisfying the condition above and the identities of \cref{prop:homomorphism}. Whenever $(\cat(X,1+1), \wedge)$ is conditionally divisible and $p\wedge\snot q\lesssim \snot(p\wedge q)$ we have:
	\begin{align*}
		(\id+\nabla)\circ (\cho(q)+\cho(q))\circ \cho(p) = (\nabla+ \id)\circ \left(\!\cho\!\left(\frac{p\wedge \snot q}{\snot(p\wedge q)}\right)\!+\cho\!\left(\frac{p\wedge \snot q}{\snot(p\wedge q)}\right)\!\right)\!\circ \cho(p\wedge q)
	\end{align*}
\end{theorem}
It is straightforward to check that the previous theorem implies that if the tests $p\wedge q$ and $\frac{p\wedge \snot q}{\snot(p\wedge q)}$ can be realised as the semantics of programs $p\ \&\ q$ and $p\ \bullet\ q$ then the following generalisation of GKAT's skew-associativity rule holds:
\begin{align*}
	&\sem{\Gamma\vdash \ite{p}{(\ite{q}{s_1}{s_2})}{s_3}} 
	\\
	= ~& \sem{\Gamma\vdash \ite{p\ \&\ q}{s_1}{(\ite{p\bullet q}{s_2}{s_3}})}.
\end{align*}

\begin{example}
	The (pointwise) boolean structure on non-deterministic tests is conditionally divisible. The relation $\lesssim$ is the standard lattice order, we trivially have that $p\wedge\snot q\lesssim \snot(p\wedge q)$, and we can take $ \frac{p\wedge \snot q}{\snot(p\wedge q)} =  ((\snot(p\wedge q)\to p\wedge \snot q)$ which is just equivalent to $p$ and recover exactly the skew-associativity of GKAT stated in \cref{sec:def}.
\end{example}

\begin{example}
	The (pointwise) involutive magma structure of probabilistic tests is conditionally divisible. The relation $\lesssim$ is the standard order on $[0,1]$, we trivially have $p(1-q)\leq (1-pq)$, and we take $ \frac{p\wedge \snot q}{\snot(p\wedge q)} =\frac{p(1-q)}{(1-pq)}$ (with the convention that $\frac{0}{0}=0$, no other zeroes in the denominator are possible by definition of conditional divisibility). We then recover the skew-associativity rule of barycentric algebras discussed  in \eg\cite{mardare2016quantitative}:
	\[
	(e+_q f)+_p g = e_{pq} (f +_{\frac{p-pq}{1-pq}} g)
	\]
\end{example}

\section{Conclusion}
We have surveyed several inequivalent semantics for conditional branching found in the literature, covering a range of computational paradigms (classical, effectful, probabilistic, quantum). We saw that even in linear (\ie resource-sensitive) semantics, such as those of probabilistic, quantum or effectful computations, it is possible to interpret non-linear \ifte statements whose guards do not consume their context, \cf typing rule \eqref{rule:cond_vanilla}. 
The variety of existing semantics begs the question: is there a unifying `minimal' structure required to interpret the \ifte rule \eqref{rule:cond_vanilla}?

To answer this question we have focused on the central construct required to interpret conditional branching, the \emph{coproduct}. Fundamentally, this is the structure that allows us to combine the interpretations of the if-branch and of the else-branch into a single program. The question is then to construct a branching behaviour into this coproduct based on a guard. This is where we introduce our notion of \emph{\name category}, a monoidal distributive category equipped with a way of turning guards -- \ie predicates -- into branching behaviours. Our definition echoes that of traced monoidal category and is `minimal' in the sense that it is as parsimonious as possible, with only two combinatorially obvious requirements enforcing the equivalences $\ite{p}{s}{t}\equiv \ite{\pnot p}{t}{s}$ and 
$\ite{\progfont{true}}{s}{t}\equiv s$. These are purely combinatorial in the sense that they refer only to the basic combinatorial properties of the coproduct (via the maps $\snot, \strue: I+I\to I+I$). Additional program equivalences follow from additional demands on the \name structure (\eg idempotence, commutativity).

We have shown that a wide range of semantically important categories possess canonical \name structures, and in particular categories supporting the two types of branching found in quantum computing: measurement-based control and coherent control. It would be nice to find a category that could cover both at the same time, either through a unifying \name structure, or through two dedicated \name structures (but defined on the same category). 

We have also shown that a \name structure gives rise to algebraic structures on sets $\cat(X, X+X)$ of branching morphisms and on sets $\cat(X, I+I)$ of predicates. The richness of these structures depends on the requirements put on the \name structure, but in most cases, we obtain an involutive magma. For non-deterministic and probabilistic computations, we recover know structures and show program equivalences for nested branching corresponding to the skew-associativity rules of GKAT and barycentric algebras respectively. We believe that a lot more can be said about the structure of predicates in \name categories, both in general and in interesting special cases.



\bibliography{conditionals.bib}

\newpage 
\appendix

\section{Proofs for Section \ref{sec:def}}
\begin{lemma}\label{lem:distr_property}
	The following diagram commutes in any distributive category
	\[
	\xymatrix{
		X\otimes (Y+Z)\ar[rr]^{f\otimes [g,h]} \ar[dr]_{\ldistr_{X,Y,Z}}& & X\otimes W
		\\
		& (X\otimes Y) +(X\otimes Z)\ar[ur]_{[f\otimes g, f\otimes h]}
	}
	\]
	and similarly for right-distributivity.
\end{lemma}
\begin{proof}
	We look at the inverse of $\ldistr_{X,Y,Z}$ and get the following commutative diagram which proves the assertion.
	\[
	\xymatrix{
		& X\otimes W
		\\
		& X\otimes (Y+Z)\ar[u]^{f\otimes [g,h]}
		\\
		X\otimes Y\ar[r]_-{\inj_1}\ar[ur]^{\id_X\otimes \inj_1\ }\ar@/^2.5pc/[uur]^{f\otimes g} & X\otimes Y+X\otimes Z\ar@{..>}[u]^-{\ldistr\inv}& X\otimes Z\ar[l]^-{\inj_2}\ar[ul]_{\id_X\otimes \inj_2} \ar@/_2.5pc/[uul]_{f\otimes h} 
	}
	\]
\end{proof}
\begin{proof}[Proof of \cref{thm:semantic_identites}]
	\begin{enumerate} 
		\item Skew-commutativity follows from the commutativity of this diagram:
		\[
		\xymatrix@C=12ex{
			\sem{\Gamma}\otimes\sem{\Delta}\ar[r]^-{\cho_{\sem{\Gamma}}(\sem{p})\otimes \id_{\sem{\Delta}}}\ar[dr]_{\cho_{\sem{\Gamma}}(\snot p)\otimes \id_{\sem{\Delta}}}
			& (\sem{\Gamma}+\sem{\Gamma})\otimes\Delta\ar[r]^{\rdistr}\ar[d]^{\snot_{\sem{\Gamma}}\otimes\id_{\sem{\Delta}}} 
			& \sem{\Gamma}\otimes\sem{\Delta}+\sem{\Gamma}\otimes\sem{\Delta}
			\ar[ddl]^{~~~\snot_{\sem{\Gamma}\otimes\sem{\Delta}}}\ar[dd]^{[\sem{s},\sem{t}]}
			\\
			& (\sem{\Gamma}+\sem{\Gamma})\otimes \sem{\Delta}\ar[d]^{\rdistr}
			\\
			&\sem{\Gamma}\otimes\sem{\Delta}+\sem{\Gamma}\otimes\sem{\Delta}\ar[r]_{[\sem{t},\sem{s}}
			&  \sem{\type}
		}
		\]
		where the left-hand triangle commutes by the skew-commutativity axiom. The bottom right-hand triangle commutes trivially and the central square can also be seen to commute by considering the inverse of $\rdistr$, namely $[\inj_1\otimes \id_{\sem{\Delta}}, \inj_2\otimes \id_{\sem{\Delta}}]$.
		\item The proof for the true test is the essentially the same of the previous case but with $\snot$ replaced by $\strue$.
		\item Idempotence follows from the commutativity of the following diagram
		\[
		\xymatrix@C=12ex{
			\sem{\Gamma}\otimes\sem{\Delta}\ar[r]^-{\cho_{\sem{\Gamma}}(\sem{p})\otimes \id_{\sem{\Delta}}}\ar@{=}[dr] & (\sem{\Gamma}+\sem{\Gamma})\otimes\Delta\ar[r]^{\rdistr}\ar[d]^{\nabla_{\sem{\Gamma}}\otimes\id_{\sem{\Delta}}} & \sem{\Gamma}\otimes\sem{\Delta}+\sem{\Gamma}\otimes\sem{\Delta}
			\ar[dl]^{~~~\nabla_{\sem{\Gamma}\otimes\sem{\Delta}}}\ar[d]^{[\sem{s},\sem{s}]}
			\\
			& \sem{\Gamma}\otimes \sem{\Delta}\ar[r]_{\sem{s}} & \sem{\type}
		}
		\]
		The left-hand triangle commutes by virtue of the idempotence axiom. The top triangle of the right-hand square commutes by \cref{lem:distr_property}, the bottom one commutes trivially.
		\item Commutativity follows immediately from the definitions.
	\end{enumerate}
\end{proof}

	\begin{proof}[Proof of \cref{prop:distributivity}]
	The morphism $ \sem{\Gamma,\Delta\vdash \ite{p}{\progfont{let}~x=s_1~\progfont{in}~t}{\progfont{let}~x=s_2~\progfont{in}~t}:\type}$ is given by the top path of the diagram below whilst the morphism $\sem{\Gamma',\Delta'\vdash \progfont{let}~x=(\ite{p}{s_1}{s_2})~\progfont{in}~t:\type}$ follows the path below it. Commutativity follows directly from the definitions.
	\[
	\xymatrix@C=30ex{
		\sem{\Gamma}\otimes\sem{\Delta}\ar@{=}[d]\ar[r]^-{\rdistr \circ \cho_{\sem{\Gamma}}(p)\otimes \id_{\sem{\Delta}}} 
		& 
		(\sem{\Gamma}\otimes\sem{\Delta})+(\sem{\Gamma}\otimes\sem{\Delta}) \ar[d]^{\left[\sem{s}_1\otimes \id_{\Delta'},\sem{s}_2\otimes \id_{\Delta'}\right]}\ar[r]^-{[\sem{\progfont{let}~x=s_1~\progfont{in}~t},\sem{\progfont{let}~x=s_2~\progfont{in}~t}]} & \sem{\type}
		\\
		\sem{\Gamma'}\otimes\sem{\Delta'}
		\ar[r]_{\sem{\ite{p}{s_1}{s_2}}}
		& \sem{\type[S]}\otimes \sem{\Delta'}\ar[ur]_-{\sem{t}}
	}
	\]
\end{proof}

	\section{Proofs for Section \ref{sec:examples}}
\begin{proof}[Proofs of \cref{thm:copydistr,thm:markov}] To simplify the notation we will often omit the right unitors in what follows and consider $X$ and $X\otimes I$ interchangeably.
	\begin{enumerate} 
		\item \emph{Skew-commutativity}:
		This identity holds by virtue of the definition of $\cho_X$ and the commutativity of the following diagram:
		\[
		\xymatrix@R=10ex@C=22ex{
			X\otimes (I+I) \ar[d]_{\id_X\otimes \snot}\ar@/_1pc/[r]^{ \ldistr_X} & X+X\ar[d]^{\snot_X=[\inj_2, \inj_1]}\ar@/_1pc/[l]_{ [\id_X\otimes \inj_1, \id_X\otimes \inj_2]}\\
			X\otimes (I+I)\ar@/_1pc/[r]^{ \ldistr_X} & X+X\ar@/_1pc/[l]_{ [\id_X\otimes \inj_1, \id_X\otimes \inj_2]}
		}
		\]
		which holds since
		\begin{align*}
			[\id_X\otimes \inj_1, \id_X\otimes \inj_2]\circ \snot_X\circ \inj_1&=	[\id_X\otimes \inj_1, \id_X\otimes \inj_2]\circ \inj_2
			\\&=  \id_X\otimes \inj_2
			\\&= (\id_X\otimes \snot) \circ (\id_X\otimes \inj_1)
			\\& =(\id_X\otimes \snot) \circ[\id_X\otimes \inj_1, \id_X\otimes \inj_2]\circ \inj_1
		\end{align*}
		and similarly for the pre-composition with $\inj_2$. The commutativity of the square with left-to-right horizontal arrows now follows by the universal property of the coproduct, and the commutativity of the square with right-to-left horizontal arrows follows from the fact that the horizontal arrows are inverse of one another.
		\item \emph{True test:} This identity can be shown to hold in the same way as the previous one. Consider the diagram
		\[
		\xymatrix@R=10ex@C=22ex{
			X\otimes (I+I) \ar[d]_{\id_X\otimes \strue}\ar@/_1pc/[r]^{ \ldistr_X} & X+X\ar[d]^{\strue_X=[\inj_1, \inj_1]}\ar@/_1pc/[l]_{ [\id_X\otimes \inj_1, \id_X\otimes \inj_2]}\\
			X\otimes (I+I)\ar@/_1pc/[r]^{ \ldistr_X} & X+X\ar@/_1pc/[l]_{ [\id_X\otimes \inj_1, \id_X\otimes \inj_2]}
		}
		\]
		We have
		\begin{align*}
			[\id_X\otimes \inj_1, \id_X\otimes \inj_2]\circ \strue_X\circ \inj_1&=	[\id_X\otimes \inj_1, \id_X\otimes \inj_2]\circ \inj_1
			\\&=  \id_X\otimes \inj_1
			\\&= (\id_X\otimes \strue) \circ (\id_X\otimes \inj_1)
			\\& =(\id_X\otimes \strue) \circ[\id_X\otimes \inj_1, \id_X\otimes \inj_2]\circ \inj_1
		\end{align*}
		and similarly for the pre-composition with $\inj_2$. The result follows as in the previous case.
		\item \emph{Commutativity:} The proof  hinges on the commutativity of the following diagram connecting distributivity, the symmetry transformations $\gamma_{A,B}: A\otimes B\to B\otimes A$ of the symmetric monoidal category, and the permutation $[\inj_1, \inj_3, \inj_2, \inj_4]$ of the commutativity definition
		\[
		\xymatrix@C=14ex{
			A\otimes C + B\otimes C + A\otimes D + B\otimes D 
			\ar[d]_{[\inj_1\otimes \inj_1, \inj_2\otimes \inj_1, \inj_2\otimes \inj_1, \inj_2\otimes\inj_2]}^{=(\rdistr +\rdistr \circ \ldistr)\inv}
			\ar[r]_{[\inj_1\circ \gamma, \inj_3\circ \gamma, \inj_2\circ \gamma, \inj_4\circ \gamma]}
			& 
			C\otimes A + D\otimes A+C\otimes B+D\otimes B\ar[d]_{(\rdistr +\rdistr \circ \ldistr)\inv}
			\\
			(A+B)\otimes (C+D)\ar[r]_{\gamma_{(A+B),(C+D)}}
			&
			(C+D)\otimes (A+ B)
		}
		\]
		The commutativity of this diagram can be checked straightforwardly by examining each summand of the top-left-hand corner individually and using the universal property of coproducts.  This establishes the commutativity of the central square in the following diagram
		\[
		{
			\hspace{-15mm}
			\footnotesize
			\xymatrix@C=2pt{
				& & & X\ar[dr]^{\cp_X}\ar[dl]_{\cp_X}
				\\
				& & X\otimes X\ar[rr]^{\gamma}\ar[d]^{q\otimes p}\ar[dl]_{\id\otimes p} & & X\otimes X\ar[d]^{p\otimes q}\ar[dr]^{\id \otimes q}
				\\
				& X\otimes (I+I)\ar[dl]_{\ldistr}\ar[r]^-{q\otimes \id }
				& (I+I)\otimes (I+I)\ar[rr]_-{\gamma}\ar[d]_{\ldistr}
				& & (I+I)\otimes (I+I)\ar[d]^{\ldistr}  
				& X\otimes (I+I)\ar[dr]^{\rdistr}\ar[l]_-{p\otimes \id }
				\\
				(X\otimes I)+(X\otimes I)\ar[rr]_{(q\otimes \id)+(q\otimes \id)}\ar[d]^{\rho_X+\rho_X} 
				& & (I+I)\otimes I+(I+I)\otimes I\ar[d]_{\rdistr +\rdistr }
				& & (I+I)\otimes I+(I+I)\otimes I \ar[d]^{ \rdistr + \rdistr }
				& & (X\otimes I)+(X\otimes I)\ar[ll]^{(p\otimes \id)+(p\otimes \id)}\ar[d]^{\rho_X+\rho_X} 
				\\
				X+X\ar[drr]_{q+q} & & I\otimes I+I\otimes I+I\otimes I+I\otimes I \ar[rr]_{[\inj_1,\inj_3,\inj_2,\inj_4]}  \ar[d]_{[\rho,\rho]+[\rho,\rho]}
				& & I\otimes I+I\otimes I+I\otimes I+I\otimes I \ar[d]^{[\rho,\rho]+[\rho,\rho]}
				& & X+X\ar[dll]^{p+p}
				\\
				& & I+I+I+I  \ar[rr]_{[\inj_1,\inj_3,\inj_2,\inj_4]}   & & I+I+I+I 
			}
		}
		\]
		The left- and right-hand pentagons can be seen to commute by showing $[\rho_I,\rho_I]\circ \rdistr = \rho_{I+I}$ using the definition of $\rdistr$ as the inverse of the canonical morphism and then applying naturality of the unitor. The result now follows immediately.
		
		\item \emph{Short-circuiting}. Assume that
		\[
			\xymatrix{
			I\ar[r]^{s}\ar[d]_{h} & X\ar[d]^{p}
			\\
			I\ar[r]_{\inj_i} & I+I
			}
		\]
		with $s$ deterministic.
		Now consider the following diagram;
		\[
			\xymatrix{
			I\ar[r]^{s}\ar[d]_{\cp_I} & X\ar[d]^{\cp_X} \ar@/_-6pc/[dddd]^{\cho_{X}(p)}
			\\
			I\otimes I\ar[r]_{s\otimes s}\ar[d]_{s\otimes h} & X\otimes X\ar[d]^{\id\otimes p}
			\\
			X\otimes I\ar[r]_{\id\otimes \inj_i} \ar[dr]_{\inj_i}\ar[dd]& X\otimes (I+I)\ar[d]^{\ldistr_X}
			\\
			& (X\otimes I)+(X\otimes I)\ar[d]
			\\
			X\ar[r]_{\inj_i}& X+X
			}
		\]
		The top square commutes by definition of deterministic morphisms. The bottom triangle is seen to commute by using the fact that $\ldistr_X$ is the inverse of $[\id_\otimes \inj_1, \id\otimes\inj_2]$.
		
		\item \emph{Idempotence}: We now assume a Markov category, in particular $I=1$ is terminal. As the following diagram shows, 
		it follows from the universal property of the coproduct that $(\id_X\otimes \nabla_1)\circ[\id_X\otimes \inj_1,\id_X\otimes \inj_2]  = \nabla_X$:
		\[
		\xymatrix{
			& X\otimes I 
			\\
			& 
			X\otimes (I+I)\ar[u]_{\id_X\otimes \nabla_1}
			\\
			X\otimes I\ar[r]_-{\inj_1} \ar@/_-1.5pc/[uur]^{\id_X\otimes \id_I=\id_{X\otimes I}}
			& 
			X\otimes I + X\otimes I \ar[u]_{[\id_X\otimes \inj_1,\id_X\otimes \inj_2]}\ar@/_-2.5pc/[uu]^{\nabla_X}
			&  
			X\otimes I\ar[l]^-{\inj_2}\ar@/^-1.5pc/[uul]_{\id_X\otimes \id_I=\id_{X\otimes I}}
		}
		\]
		We now have (ignoring the trivial role of the right unitor $X\otimes I\simeq X$)
		\begin{align*}
			& \nabla_X\circ \cho_X(p) \\
			=~&(\id_X\otimes  \nabla_1)\circ[\id_X\otimes \inj_1,\id_X\otimes \inj_2] \circ \cho_X(p) \\
			=~& (\id_X\otimes \nabla_1)\circ[\id_X\otimes \inj_1,\id_X\otimes \inj_2] \circ \ldistr_X\circ (\id_X\otimes p)\circ  (\cp_X\otimes \id_I)\\
			=~& (\id_X\otimes \nabla_1)\circ(\id_X\otimes p)\circ  (\cp_X\otimes \id_I) \\
			=~& (\id_X\otimes !_{X\otimes I})\circ  (\cp_X\otimes \id_I)\\
			\simeq ~& \id_{X\otimes I}
		\end{align*}
		where the penultimate step follows from the fact that $I$ is terminal (and thus $\nabla_1\circ p=!_{X\otimes I}$) and the last step follows from the copy-discard interaction (up to a unitor, hence the $\simeq$).

	\end{enumerate}
\end{proof}

\begin{proof}[Proof of \ \cref{prop:cdcat_shortcrcuit}]
	Let us first show that if a map $I\stackrel{f}{\longrightarrow} X+X\stackrel{\del+\del}{\longrightarrow} I+I $ factors through $\inj_1: I\to I+I$ then $f$ factors through $\inj_1: X\to X+X$. By the weak pullback assumption we have a unique mediating morphism
	\[
	\xymatrix@C=10ex@R=4ex{
		I\ar@/^1pc/[drr]^{f}\ar@/_1pc/[ddr]_{g}\ar@{-->}[dr]
		\\
		& X\ar[r]^{\inj_1} \ar[d]_{\del}
		& X+X\ar[d]^{\del+\del}
		\\
		& I\ar[r]_{\inj_1}
		& I+I
	}
	\]
	and similarly if $(\del+\del)\circ f=\inj_2$. Next we show that given a predicate $p:X\to I+I$ and a any state $s: I\to X$, if $p\circ s$ factors through $\inj_1: I\to I+I$, then so does
	\[
	(\del+\del)\circ \cho_X(p)\circ s
	\]	
	and similarly for $\inj_2$. This follows immediately from the commutativity of the diagram:
	\[
	\xymatrix@C=8ex{
		I\ar[d]_{s}
		\\
		X \ar@/^3pc/[rrrr]_{\cho_X(p)}
		\ar@/_1.5pc/[ddrrr]_{p}\ar[r]^{\cp_X}
		& X\otimes X \ar@/_0.5pc/[dr]_-{\del\otimes p}\ar[r]^{\id\otimes p}
		& X\otimes (I+I) \ar[r]^-{\ldistr}\ar[d]_{\del\otimes (\id_I+\id_I)}
		& (X\otimes I)+(X\otimes I)\ar[d]_-{[\del\otimes\id_I,\del\otimes\id_I]} \ar[r]^-{\rho_X+\rho_X}
		& X+X\ar[ddl]^{\del+\del}
		\\
		& & I\otimes (I+I)\ar[r]_{\ldistr}  & (I\otimes I)+(I\otimes I)\ar[d]_{\rho_I+\rho_I}
		\\
		& & &   I+I
	}
	\]
	Note that the CD-category structure is essential for the diagram above to commute
\end{proof}

\begin{proof}[Proof of \cref{thm:pure_tests}]
	The proof follows from the fact that the copairing of $f: X\to Z, g: Y\to Z$ in $\Kl(T)$ is given by $\mu\circ T[f,g]$ in the base category, and the commutativity of the following diagrams:
	\begin{enumerate}
		\item \emph{Skew-commutativity:}
		\[
		\xymatrix@C=12ex{
			X \ar@/^1.5pc/[rrr]^{\widehat{\cho}_X(p)}\ar[ddr]_{\cho_X(\snot p)}\ar[r]_-{\cho_X(p)}  & X+X\ar[dr]^{\qquad[T\inj_2\circ \eta_X, T\inj_1\circ \eta_X]}\ar[dd]^>>>>>>{\snot_X=[\inj_2,\inj_1]}\ar[rr]^{\eta_X} & & T(X+X)\ar[d]^{T[T\inj_2\circ \eta_X,T\inj_1\circ \eta_X]}
			\\
			& & T(X+X)\ar@{=}[dr]]\ar[r]_{\eta_{T(X+X)}} & T^2(X+X)\ar[d]^{\mu_{X+X}}
			\\
			& X+X\ar[rr]_{\eta_X} & & T(X+X)
		}
		\]
		The bottom-left triangle commutes by naturality of $\eta_X$ applied to both component of the copairing. Again the right-hand-vertical path is the copairing $[\inj_2, \inj_1]$, that is to say $\snot_X$, in the category $\Kl{T}$.
		\item \emph{True test:} The proof is the same as for skew-commutativity with $\snot$ replaced by $\strue$.
		\item \emph{Idempotence:}
		\[
		\xymatrix@C=12ex{
			X\ar@/^1.5pc/[rrr]^{\widehat{\cho}_X(p)}\ar@{=}[ddr]\ar[r]_-{\cho_X(p)} & X+X\ar[dd]_{\nabla_X}\ar[rr]_-{\eta_{X+X}}\ar[dr]^{[\eta_X,\eta_X]} & & T(X+X) \ar[d]^{T[\eta_X,\eta_X]}
			\\
			& & TX\ar[r]^{\eta_{TX}}\ar@{=}[dr] & T^2X\ar[d]^{\mu_X}
			\\
			& X\ar[rr]_{\eta_X} & & TX
		}
		\] 
		Where the right-hand-side vertical path is the codiagonal morphism in $\Kl{T}$  and the bottom path the unit in $\Kl{T}$.
	\end{enumerate}
\end{proof}

For the following proof we must assume some familiarity with the theory of Riesz spaces, Banach lattices and positive operators, see~\cite[Chapters 8,9]{aliprantis2006infinite} for a gentle introduction and~\cite{aliprantis2006positive} or \cite{zaanen2012introduction} for a more in-depth account.
\begin{proof}[Proof of \cref{thm:vector_split}]
	Let us write $ev_p$ for the functional $\nabla_{\R}\circ p$ (an evaluation functional in the probabilistic case), $p_t$ for $\pi_1\circ p$ (which measures how true the guard is for a given vector) and $p_f$ for $\pi_2\circ p$.
	If $p_t=ev_p$ (and thus $p_f=0$) then take $x=x_t$ and $x_f=0$ and conversely if $p_f=ev_p$.
	
	Now assume that $p_t\defeq \pi_1\circ p$ and $p_f\defeq \pi_2\circ p$ are strictly positive and proceed in steps.
	\begin{itemize}
		\item \textbf{$X$ is Archimedean and has the countable sup property}: See~\cite[Thm 8.22]{aliprantis2006infinite}
		\item \textbf{$X$ is Dedekind complete}: This follows immediately from the fact that it is $\sigma$-Dedekind complete and has the countable sup property.
		\item \textbf{$X$ has the Principal Projection Property}: Let $x\in X$ and let $B_x$ be the principal band generated by $x$. For $0\leq y$ we define the projection $P_x: X\to B_x$ as follows:
		\[
		P_x(y) = \bigvee\{0\leq u\leq y\mid u\in B_x\}.
		\]
		Since $0$ is certainly in the set above, and since it is bounded from above by $y$, the sup above exists by Dedekind-completeness. Since $B_x$ is a band, it is closed under joins of directed sets and $\{0\leq u\leq y\mid u\in B_x\}$ is certainly a directed set of elements of $B_x$. It follows that $P_x(y)\in B_x$ and thus that $P_x^2=P_x$, \ie $P_x$ is a projection. 
		
		To see that $P_x$ is linear let $y,z\in X$ and note first that since $P_x(y), P_x(z)\in B_x$ then $P_x(y)+P_x(z)\in B_x$, moreover $P_x(y)\leq y$ and $P_x(z)\leq z$ thus $P_x(y)+P_x(z)\leq y+z$ and thus $P_x(y)+P_x(z)\leq P_x(y+z)$. For the other direction we use the fact that $X$ is necessarily a Riesz space (since $\{u,v\}$ is bounded above by $u+v$ it follows that $\bigvee\{u,v\}$ necessarily exists) and therefore has the Riesz decomposition property: for every $0\leq w\leq y+z$, there exists $0\leq w_1\leq y, 0\leq w_2\leq z$ such that $w=w_1+w_2$. It follows that for any element $w$ in the set defining $P_x(y+z)$ we can find a pair of elements $w_1, w_2$ in the sets defining $P_x(y)$ and $P_x(z)$ in particular $w_1+w_2\leq P_x(y)+P_x(z)$. Since this holds for every element in the set defining $P_x(y+z)$ we get $P_x(y+z)\leq P_x(y)+P_x(z)$. 
		
		Finally, we need to show that $y-P_x(y)$ is disjoint from $P_x(y)$. If $0< (y-P_x(y)\wedge y)\leq P_x(y)$ then $0<(y-P_x(y)\wedge y)\in B_x$ and thus $(y-P_x(y)\wedge y) + P_x(y)\in B_x$. Since $(y-P_x(y)\wedge y) + P_x(y) \leq y-P_x(y)+P_x(y)=P_x(y)\leq y$ it belongs to the set defining $P_x(y)$ but also $P_x(y)< (y-P_x(y)\wedge y) +P_x(y) $ which contradicts the definition of $P_x(y)$ as the supremum of such elements.
		
		\textbf{Construction of $x_t$ and $x_f$}.  Note first that $p_t\leq ev_p$ and therefore $p_t$ belongs to the ideal $E_{ev_p}$ generated by $ev_p$ in $X^\sim$, the space of order-bounded linear functionals on $X$. Since $\R$ is a Dedekind-complete  Riesz space, $X^\sim$ is a Dedekind-complete Riesz space \cite[Thm.1.18]{aliprantis2006positive}, in particular it has the principal projection property as shown above. We therefore have all the conditions required to apply Freudenthal's Spectral Theorem \cite[Thm.2.8]{aliprantis2006positive} to $p_t\in E_{ev_p}$. This guarantees the existence of an increasing sequence $\xi_n\uparrow p_t$ of $ev_p$-simple functions, that is to say linear combinations of components of $ev_p$
		\[
		\xi_n=\sum_{i}^{N_n}\alpha_i \xi_{n,i}
		\]
		where $\sum_{i}\xi_{n,i}=ev_p$ and $\xi_{n,i}\wedge\xi_{n,j}=0, i\neq j$. Since $X$ is Dedekind complete we can define for every $\xi_{n,i}$ the vector
		\[
		y_{n,i}=\bigwedge\{0\leq u\leq x\mid \xi_{n,i}(u)=\xi_{n,i}(x)\}
		\]
		We now associate to $\xi_n$ the vector
		\[
		y_n = \sum_{i}^{N_n}\alpha_i y_{n,i}\in X.
		\]
		We claim that $y_n\leq x$. If this is the case, and since $X$ is $\sigma$-Dedekind complete we can define
		\[
		x_t \defeq \bigvee_n y_n.
		\]
		The null ideal of $\xi_{n,i}$ is defined as the set $N_{n,i}=\{x\in X\mid \xi_{n,i}(\abs{x})=0\}$. Its disjoint complemented $N_{n,i}^d=\{x\in X\mid \abs{x}\perp \abs{y}\text{ for all }y\in N_{n,i}\}$ is called the carrier $C_{n,i}$ of $\xi_{n,i}$. Like the disjoint complement of any subset it is a band \cite[\S 8.9]{aliprantis2006infinite} and since every band is a projection band in a Dedekind complete Riesz \cite[Thm.1.41]{aliprantis2006positive} $X=C_{n,1}\oplus C_{n,1}^d$. We can then decompose $C_{n,1}^d$ according to $\xi_{n,2}$ in the same way and decompose the entire space $X$ into bands
		\[
		X= C_{n,1}\oplus \ldots \oplus C_{n,N_n}\oplus B_{rest}
		\]
		since $\xi_{n,i}\perp\xi_{n,j}$ is equivalent to $C_{n,i}\perp C_{n,j}$ \cite[Thm.1.67]{aliprantis2006positive}. Thus for any $x\in X$ we can find $0\leq x_i\in C_{n,i}$ such that 
		$x_1+\ldots+x_{N_n}\leq x$ and $\xi_{n,i}(x_i)=\xi_{n,i}(x)$. Clearly $y_{n,i}\leq x_i$.
		
		Since $\xi_n=\sum_{i}\alpha_i\xi_{n,i}\leq \sum_{i}\xi_{n,i}=ev_p$ and the $\xi_{n,i}$ are pairwise disjoint ($\xi_{n,i}\perp\xi_{n,i}$), it follows that each $\alpha_i\leq 1$. We therefore have
		\[
		y_n=\sum_{i}\alpha_i y_{n,i}\leq \sum_{i}y_{n,i}\leq \sum_{i}x_i\leq x
		\]
		as claimed, i.e.\ $x_t\in B_x$. 
		
		We now define $x_f \defeq x-x_t\in B_x$ which proves the first two assertions.
		
		Now we need to show that the assignment $x\mapsto (x_t, x_f)$ is linear (it is clearly positive). To do this we just need to check that $x\mapsto x_t$ is linear. To define $x_t+x'_t$ at stage $n$ we consider the sets 
		\[
		\{0\leq u\leq x\mid \xi_{n,i}(u)=\xi_{n,i}(x)\}\qquad\{0\leq u'\leq x\mid \xi_{n,i}(u')=\xi_{n,i}(x')\}
		\]
		For any such $u,u'$ $\xi_{n,i}(u+u')\leq \xi_{n,i}(x+x')$ and thus these $u,u'$ are elements in 
		\[
		\{0\leq u\leq x\mid \xi_{n,i}(u)=\xi_{n,i}(x+x')\}
		\]
		and it follows that 
		\begin{align*}
			&\bigwedge\{0\leq u\leq x\mid \xi_{n,i}(u)=\xi_{n,i}(x+x')\}\leq \\
			&\bigwedge\{0\leq u\leq x\mid \xi_{n,i}(u)=\xi_{n,i}(x)\}+\bigwedge\{0\leq u'\leq x\mid \xi_{n,i}(u')=\xi_{n,i}(x')\} =\\
			&\bigwedge\{0\leq u+u'\leq x+x'\mid \xi_{n,i}(u+u')\leq \xi_{n,i}(x+x')\}
		\end{align*}
		For the converse we use the Riesz decomposition property as above. Thus the sequence defining $(x+x')_t$ is constructed as the sum of the sequences defining $x_t, x'_t$ and thus
		\[
		(x+x')_t = \bigvee_n (y_n+y'_n)=\bigvee_n y_n + \bigvee_n y_n'=x_t+x'_t.
		\]
		
		\item \textbf{Case of AL-spaces}: Every AL-space has an order continuous norm \cite[\S9.5]{aliprantis2006infinite} and thus
		\begin{align*}
			\norm{x_t}&=\norm{\bigvee_n y_n}\\
			&=\bigvee_n\norm{\sum_i \alpha_i y_{n,i}}&\text{Norm is order-continuous}
			\\
			&=\bigvee_n \sum_i\alpha_i\norm{y_{n,i}}&\text{AL-space}
			\\
			&=\bigvee_n \sum_i\alpha_i ev_p(y_{n,i})&ev_p\text{ norm-preserving}
			\\
			&=\bigvee_n \sum_i\alpha_i \xi_{n,i}(y_{n,i})&y_{n,i}\in C_{n,i}
			\\
			&=\bigvee_n \sum_i\alpha_i \xi_{n,i}(x)&\text{definition of }y_{n,i},\xi_{n,i}\text{ order-continuous}
			\\
			&=\bigvee_n \xi_n(x)
			\\
			&=p_t(x)
		\end{align*}
		
		\item \textbf{\Name structure}. The skew-commutativity, true test axioms, idempotence and short-circuiting property follow directly from the definition of $x_f$ as $x-x_t$ and $x_t=x$ when $p_f=ev_p$.
	\end{itemize}
\end{proof}

\begin{proof}[Proof of \cref{prop:quantum_split}] Both the $\Hilb$ and $\CPMp$ case can be treated simultaneously by reasoning directly on the if and else operators $I_p$ and $E_p$.
	\begin{enumerate}
		\item Skew-associativity: We show that $I_{\snot p}=E_p$ and the result follows immediately. Note that $\snot: \C^2\to\C^2$ is given by 
		$\begin{pmatrix}
			0 & 1\\1 & 0
		\end{pmatrix}$,
		$\pi_1: \C^2\to\C^2$ by 
		$\begin{pmatrix}
			1 & 0\\0 & 0
		\end{pmatrix}$,
		and $\pi_2: \C^2\to\C^2$ by 
		$\begin{pmatrix}
			0 & 0\\0 & 1
		\end{pmatrix}$
		in the computational basis.
		\begin{align*}
			I_{\snot p} & \defeq (\snot p)\dg \pi_1 (\snot p)\\
			&=p\dg \snot\dg \pi_1 \snot p\\
			&=p\dg\pi_2 p\\
			&=E_p
 		\end{align*}
	 	\item True test: we show that $I_{\strue p}=\id_p$ and $E_{\strue p}=0$ and the result then follows from the fact that $\strue\circ (E_p,I_p)^T=\begin{pmatrix}
	 		1 & 1\\0 & 0
	 	\end{pmatrix}\begin{pmatrix}
	 	E_p\\I_p
	 	\end{pmatrix}=\begin{pmatrix}E_p+I_p\\0\end{pmatrix}=\begin{pmatrix}\id_X\\0\end{pmatrix}$.
	 	\begin{align*}
	 		I_{\strue p} & \defeq (\strue p)\dg \pi_1 (\strue p)\\
	 		&=p\dg \strue\dg \pi_1 \strue p\\
	 		&=p\dg\begin{pmatrix}
	 			1 & 1\\1 & 1
	 		\end{pmatrix}p\\
	 		&= \id_X
	 	\end{align*}
	 	Conversely,
	 	\begin{align*}
	 		E_{\strue p} & \defeq (\strue p)\dg \pi_2 (\strue p)\\
	 		&=p\dg \strue\dg \pi_2 \strue p\\
	 		&=0
	 	\end{align*}
		\item Idempotence for sharp predicates follows immediately from the fact that $I_p, E_p$ are then orthogonal projections, and $E_p=\id-I_p$. The result then follows from the fact that the codiagonal map sums the two components.
		\item Commutativity is a trivial combinatorial consequence of the choice function definitions in terms of $I_p, E_p$.
	\end{enumerate}
\end{proof}

\section{Proofs for Section \ref{sec:structure}}

	The following lemma is a direct consequence of the enriched Yoneda lemma~\cite[\S1.9]{kelly1982basic}.
	\begin{lemma}\label{lem:natT}
		Let $(\cat, \otimes, I)$ be a closed symmetric monoidal category with coproducts which is enriched over itself and such that $\Id = \cat(I,-)$, and let $T=\Id+\Id$. The set of natural transformations $\Id\to T$ is in bijective correspondence with $\cat(I,I+I)$. The set of natural transformations $T^2\to T$ is in bijective correspondence with $\left(\cat(I,I+I)\right)^4$.
	\end{lemma}

	\begin{proof}[Proof of \cref{thm:monad_struct_set}]
	It follows from \cref{lem:natT} that there are only two possible units for $T$ given respectively by $\eta=\inj_1$ and $\eta=\inj_2$. Let's consider each of these options in turn.
	
	\noindent\emph{Case $\eta=\inj_1$:} again from \cref{lem:natT} there exists 16 different possible multiplication $\mu: T^2\to T$ that are all of the shape
	\[
	[\inj_{i_1},\inj_{i_2},\inj_{i_3},\inj_{i_4}]\qquad\text{with}\qquad i_1,i_2,i_3,i_4\in\{1,2\}
	\] 
	The unit axiom $\mu\circ \eta_T=\Id_T$ implies that $i_1=1$ and $i_2=2$. The other unit axiom $\mu\circ T\eta=\id_T$ implies that $i_1=1$ and $i_3=2$. This leaves two options open. The associativity axiom $\mu\circ \mu_{T}=\mu\circ T\mu$ looks like this
	\[
	\xymatrix@C=16ex{
		((X+X)+(X+X))+((X+X)+(X+X))\ar[d]_{[\inj_1,\inj_2,\inj_{i_3},\inj_{i_4}]+[\inj_1,\inj_2,\inj_{i_3},\inj_{i_4}]} \ar[r]^-{[\inj_1,\inj_2,\inj_{i_3},\inj_{i_4}]} & (X+X)+(X+X)\ar[d]^{[\inj_1,\inj_2,\inj_{i_3},\inj_{i_4}]}\\
		(X+X)+(X+X)\ar[r]_{[\inj_1,\inj_2,\inj_{i_3},\inj_{i_4}]} & X+X
	}
	\]
	Let's consider the two different options.
	\begin{enumerate}
		\item $i_4=1$. Clockwise we get the map $[\inj_1, \inj_2, \inj_2, \inj_1,\inj_2, \inj_1,\inj_1, \inj_2]$. Anti-clockwise we get the map $[\inj_1, \inj_2,\inj_2,\inj_1,\inj_2,\inj_1,\inj_1,\inj_2]$. These are the same maps, and thus this option is possible.
		\item $i_4=2$. Clockwise we get the map $[\inj_1, \inj_2, \inj_2, \inj_2,\inj_2, \inj_2,\inj_2, \inj_2]$. Anti-clockwise we also get the map $[\inj_1, \inj_2,\inj_2,\inj_2,\inj_2,\inj_2,\inj_2,\inj_2]$, and  so this option is possible too. 
	\end{enumerate}
	
	\noindent\emph{Case $\eta=\inj_2$:}  the proof is dual to the case where $\eta=\inj_1$.
\end{proof}

\begin{proof}[Proof of \cref{prop:kleisli_structure_tests}]
	\begin{enumerate}
		\item To see that $\top\vee \cho(p)=\cho(p)\vee\top=\top$ holds, consider first the following two commutative diagrams:
		\[
		\xymatrix@C=8ex{
			X\ar[r]^{\top=\inj_1}\ar[d]_{\cho(p)} & X+X\ar[r]^-{\cho(p)+\cho(p)} & X+X+X+X\ar[d]^{\vee}
			\\
			X+X\ar[urr]_{\inj_1+\inj_1}\ar[rr]_{\strue} & & X+X
		}
		\qquad\qquad
		\xymatrix@C=8ex{
			X\ar[r]^-{\cho(p)} & X+X\ar[r]^-{\inj_1+\inj_1}\ar[dr]_{\strue} & X+X+X+X\ar[d]^{\wedge}
			\\
			& & X+X
		}
		\]
		Now note that thanks to idempotency we have
		\[
		\xymatrix{
			X\ar@{=}[dr]\ar[r]^{\cho(p)} & X+X\ar[r]^{\strue} \ar[d]^{\nabla}& X+X
			\\
			& X\ar[ur]_{\top=\inj_1}
		}
		\]
		\item The proof of $\bot\wedge \cho(p)=\cho(p)\wedge\bot=\bot$ is dual to the previous case.
		\item For the identity $\snot(\cho(p)\wedge\cho(q))=\cho(\snot p)\vee \cho(\snot q)$ consider the following diagram
		\[
		\xymatrix@C=13ex{
			X\ar[r]^-{\cho(p)}\ar[dr]_{\cho(\snot p)} & X+X\ar[r]^-{\cho(q)+\cho(q)}\ar[d]^{\snot} & X+X+X+X\ar[r]^-{\wedge}\ar[d]^{\snot} & X+X\ar[dd]^{\snot}
			\\
			& X+X\ar[r]^-{\cho(q)+\cho(q)}\ar[dr]_-{\cho(\snot q)+\cho(\snot q)\qquad} & X+X+X+X\ar[d]^{\snot+\snot}
			\\
			& & X+X+X+X\ar[r]_-{\vee} & X+X
		}
		\]
		To see that the right-hand side square commutes we simply compute the maps in both directions. Clockwise the map is $[\inj_2, \inj_1, \inj_1, \inj_1]$, anticlockwise it is given by $\vee\circ [\inj_4, \inj_3, \inj_2, \inj_1]$ which is also equal to $[\inj_2, \inj_1, \inj_1, \inj_1]$.
		\item The proof of $\snot(\cho(p)\vee\cho(q))=\cho(\snot p)\wedge \cho(\snot q)$ is dual to the one above.
		\[
		\xymatrix@C=13ex{
			X\ar[r]^-{\cho(p)}\ar[dr]_{\cho(\snot p)} & X+X\ar[r]^-{\cho(q)+\cho(q)}\ar[d]^{\snot} & X+X+X+X\ar[r]^-{\vee}\ar[d]^{\snot} & X+X\ar[dd]^{\snot}
			\\
			& X+X\ar[r]^-{\cho(q)+\cho(q)}\ar[dr]_-{\cho(\snot q)+\cho(\snot q)\qquad} & X+X+X+X\ar[d]^{\snot+\snot}
			\\
			& & X+X+X+X\ar[r]_-{\wedge} & X+X
		}
		\]
		Again, the commutativity of the right-hand square follows by computing both morphisms which are equal to $[\inj_2, \inj_2, \inj_2, \inj_1]$.
		\item The commutativity $\cho(p)\vee \cho(q) = \cho(q)\vee \cho(p)$ follows immediately from the fact that $\vee = \vee\circ [\inj_1, \inj_3,\inj_2,\inj_4]$.
		\item Similarly, the commutativity $\cho(p)\wedge \cho(q) = \cho(q)\wedge \cho(p)$ follows from the fact that $\wedge = \wedge \circ [\inj_1, \inj_3,\inj_2,\inj_4]$.
	\end{enumerate}
\end{proof}

\begin{proof}[Proof of \cref{prop:commutative}]
	The de Morgan relations follow from the fact that $\cho_{X}(\snot p)=\snot_X\circ(p)$ holds in any \name category and diagram chasing around permutations of coproducts.
	For commutativity the proof follows immediately from noticing that 
	\[
	\vee = \vee\circ  [\inj_1, \inj_3, \inj_2, \inj_4] : I+I+I+I\to I+I
	\]
	and similarly for $\wedge$.
\end{proof}

\begin{proof}[Proof of \cref{prop:homomorphism}] For notational convenience we use a \name structure defined by \eqref{eq:copydistr} in the special case where $\otimes$ is the Cartesian product. The general proof can then be obtained by making all copying explicit and replacing all instances of $\Delta$ below with $\cp$, although this is laborious for pairings. Consider the following diagram which encodes the property for $\vee$ (the proof for $\wedge$ is dual).
	\[
	\hspace{-12mm}
	\xymatrix@C=2ex{
		X\ar[dd]_-{\langle \id, p\rangle}\ar[rrr]^{\langle\id,p\vee q\rangle}\ar[dr]^{\langle\id,\langle\id,p\rangle\rangle} & & & X\times (1+1)\ar[rr]^-{\ldistr} & & X+X
		\\
		& X\times X\times (1+1) \ar[dr]^{\id\times  \ldistr}\ar[ddr]_{\ldistr}
		\\
		X\times (1+1)\ar[dr]_{\ldistr}\ar[ur]_{\Delta\times (\id+\id)}& & X\times (X+X)\ar[dr]^{\id\times (q+q)}
		\ar[d]^{\ldistr}
		\\
		&  X+X\ar[ddrr]_{\langle\id, q\rangle+\langle\id, q\rangle}\ar[r]^-{\Delta+\Delta}& 
		X\times X+X\times X\ar[ddr]^{\id\times q+\id\times q}
		&  X\times (1+1+1+1)\ar[uuu]^{\id\times \vee}\ar[dd]^{\ldistr}
		\\
		\\
		& & & X\times (1+1) + X\times (1+1) \ar[rr]_-{\ldistr+\ldistr} \ar@/_5pc/[uuuuu]_{[\id\times \strue, \id\times \id]}& & X+X+X+X\ar[uuuuu]^{\vee}
	}
	\]
	The top triangle with base $\langle \id, p\vee q\rangle$ commute by definition of $p\vee q$. The left-most triangle commutes trivially and the square to the bottom-right of it commutes by naturality of $\ldistr$ as does the other square to its right involving two $\ldistr$-labelled vertical arrows. The lowest triangle involving $\Delta+\Delta$ commutes trivially and the half-moon diagram commutes by \cref{lem:distr_property}. The central triangle involving only $\ldistr$ also commutes by virtue of \cref{lem:distr_property} by considering the inverses of the two diagonal edges (the inverse of the long edge is $[\id_X\times (\id_X\times \inj_1), \id_X \times (\id_X\times \inj_2)]$). Finally we can check explicitly that the right-most square (with the bendy vertical arrow) commutes by again considering the inverse of $\ldistr$. Starting from the first summand of the bottom-right-hand object we get
	\begin{align*}
		(\id_X\times \strue) \circ\ldistr\inv &=
		(\id_X\times \strue) \circ [\id_X\times \inj_1, \id_X\times \inj_2] \\
		&=[\id_X\circ \inj_1, \id_X\circ \inj_1]
		\\
		& = [\id_X\circ \inj_1, \id_X\circ \inj_2]\circ \strue_X
		\\
		&=\ldistr\inv \circ \strue_X
	\end{align*}
	The identity for the second summand is trivial and the commutativity of the final square follows. 
\end{proof}

\begin{proof}[Proof of \cref{thm:skew-assoc}]
	We prove the equality of these morphisms by showing that they are equal when post-composed with the jointly monic arrows $[\id_X,\nabla_X]$ and $[\snot_X,\id_X]$. In the following diagram we write $r$ for the test $\frac{p\wedge \snot q}{\snot(p\wedge q)}$. That it commutes is a direct consequence of the fact that $\cat(X, 1+1)$ is residuated.
	\[
	\hspace{-8mm}
	{\small
	\xymatrix{
		X \ar[rr]^{\cho(p)}\ar[dd]_{\cho(p\wedge q)}& & X+X\ar[rr]^-{\cho(q)+\cho(q))} \ar[dr]_{\cho(\snot q)+\cho(\snot(q))~~~~}& & X+X+X+X\ar[dl]^{\snot+\snot}\ar[dddd]^{\id+\nabla}
		\\
		& & & X+X+X+X\ar[dl]^{\wedge}
		\\
		X+X\ar[r]^{\snot}\ar[dd]_{\cho(r)+\cho(r)} & X+X\ar[d]_{\cho(r)+\cho(r)}  & X+X
		\\
		& X+X+X+X\ar[ur]_{\wedge}
		\\
		X+X+X+X\ar[ur]_{\snot }\ar[rrrr]_{\nabla+\id} & & & & X+X+X\ar[lluu]^{[\snot,\id]}
	}
	}
	\]
	The second equality follows from the following commutative diagram which follows from idempotence and the definition of $\wedge$.
	\[
	\xymatrix@C=10ex{
		X\ar[r]^{\cho(p)}\ar[d]_{\cho(p\wedge q)} & X+X\ar[r]^-{\cho(q)+\cho(q)} & X+X+X+X\ar[dd]^{\id+\nabla}\ar[dl]^{\wedge}
		\\
		X+X\ar@{=}[r]\ar[d]_{\cho(r)+\cho(r)} & X+X
		\\
		X+X+X+X\ar[rr]_{\nabla+\id} & & X+X+X\ar[ul]_{[\id,\nabla]}
	}
	\]
\end{proof}

\end{document}